

\magnification=\magstep1

\catcode`\@=11
\font\tensmc=cmcsc10      
\def\smc{\tensmc}

\def\hcorrection#1{\advance\hoffset by #1 }
\def\vcorrection#1{\advance\voffset by #1 }
\def\wlog#1{}
\newif\iftitle@
\outer\def\title{\title@true\vglue 24\p@ plus 12\p@ minus 12\p@
   \bgroup\let\\=\cr\tabskip\centering
   \halign to \hsize\bgroup\tenbf\hfill\ignorespaces##\unskip\hfill\cr}
\def\endtitle{\cr\egroup\egroup\vglue 18\p@ plus 12\p@ minus 6\p@}
\outer\def\author{\iftitle@\vglue -18\p@ plus -12\p@ minus -6\p@\fi\vglue
    12\p@ plus 6\p@ minus 3\p@\bgroup\let\\=\cr\tabskip\centering
    \halign to \hsize\bgroup\smc\hfill\ignorespaces##\unskip\hfill\cr}
\def\endauthor{\cr\egroup\egroup\vglue 18\p@ plus 12\p@ minus 6\p@}
\outer\def\heading{\bigbreak\bgroup\let\\=\cr\tabskip\centering
    \halign to \hsize\bgroup\smc\hfill\ignorespaces##\unskip\hfill\cr}
\def\endheading{\cr\egroup\egroup\nobreak\medskip}

\outer\def\endproclaim{\par\ifdim\lastskip<\medskipamount\removelastskip
  \penalty 55 \fi\medskip\rm}
\outer\def\demo#1{\par\ifdim\lastskip<\smallskipamount\removelastskip
    \smallskip\fi\noindent{\smc\ignorespaces#1\unskip:\enspace}\rm
      \ignorespaces}

\newcount\footmarkcount@
\footmarkcount@=1
\def\makefootnote@#1#2{\insert\footins{\interlinepenalty=100
  \splittopskip=\ht\strutbox \splitmaxdepth=\dp\strutbox
  \floatingpenalty=\@MM
  \leftskip=\z@\rightskip=\z@\spaceskip=\z@\xspaceskip=\z@
  \noindent{#1}\footstrut\rm\ignorespaces #2\strut}}
\def\footnote{\let\@sf=\empty\ifhmode\edef\@sf{\spacefactor
   =\the\spacefactor}\/\fi\futurelet\next\footnote@}
\def\footnote@{\ifx"\next\let\next\footnote@@\else
    \let\next\footnote@@@\fi\next}
\def\footnote@@"#1"#2{#1\@sf\relax\makefootnote@{#1}{#2}}
\def\footnote@@@#1{$^{\number\footmarkcount@}$\makefootnote@
   {$^{\number\footmarkcount@}$}{#1}\global\advance\footmarkcount@ by 1 }

\hyphenation{man-u-script man-u-scripts ap-pen-dix ap-pen-di-ces}
\hyphenation{data-base data-bases}
\ifx\amstexloaded@\relax\catcode`\@=13
  \endinput\else\let\amstexloaded@=\relax\fi
\newlinechar=`\^^J
\def\eat@#1{}
\def\Space@.{\futurelet\Space@\relax}
\Space@. %
\newhelp\athelp@
{Only certain combinations beginning with @ make sense to me.^^J
Perhaps you wanted \string\@\space for a printed @?^^J
I've ignored the character or group after @.}
\def\futureletnextat@{\futurelet\next\at@}
{\catcode`\@=\active
\lccode`\Z=`\@ \lowercase
{\gdef@{\expandafter\csname futureletnextatZ\endcsname}
\expandafter\gdef\csname atZ\endcsname
   {\ifcat\noexpand\next a\def\next{\csname atZZ\endcsname}\else
   \ifcat\noexpand\next0\def\next{\csname atZZ\endcsname}\else
    \def\next{\csname atZZZ\endcsname}\fi\fi\next}
\expandafter\gdef\csname atZZ\endcsname#1{\expandafter
   \ifx\csname #1Zat\endcsname\relax\def\next
     {\errhelp\expandafter=\csname athelpZ\endcsname
      \errmessage{Invalid use of \string@}}\else
       \def\next{\csname #1Zat\endcsname}\fi\next}
\expandafter\gdef\csname atZZZ\endcsname#1{\errhelp
    \expandafter=\csname athelpZ\endcsname
      \errmessage{Invalid use of \string@}}}}
\def\atdef@#1{\expandafter\def\csname #1@at\endcsname}
\newhelp\defahelp@{If you typed \string\define\space cs instead of
\string\define\string\cs\space^^J
I've substituted an inaccessible control sequence so that your^^J
definition will be completed without mixing me up too badly.^^J
If you typed \string\define{\string\cs} the inaccessible control sequence^^J
was defined to be \string\cs, and the rest of your^^J
definition appears as input.}
\newhelp\defbhelp@{I've ignored your definition, because it might^^J
conflict with other uses that are important to me.}
\def\define{\futurelet\next\define@}
\def\define@{\ifcat\noexpand\next\relax
  \def\next{\define@@}%
  \else\errhelp=\defahelp@
  \errmessage{\string\define\space must be followed by a control
     sequence}\def\next{\def\garbage@}\fi\next}
\def\undefined@{}
\def\preloaded@{}
\def\define@@#1{\ifx#1\relax\errhelp=\defbhelp@
   \errmessage{\string#1\space is already defined}\def\next{\def\garbage@}%
   \else\expandafter\ifx\csname\expandafter\eat@\string
         #1@\endcsname\undefined@\errhelp=\defbhelp@
   \errmessage{\string#1\space can't be defined}\def\next{\def\garbage@}%
   \else\expandafter\ifx\csname\expandafter\eat@\string#1\endcsname\relax
     \def\next{\def#1}\else\errhelp=\defbhelp@
     \errmessage{\string#1\space is already defined}\def\next{\def\garbage@}%
      \fi\fi\fi\next}
\def\famzero{\fam\z@}

\def\dim{\mathop{\famzero dim}\nolimits}
\def\exp{\mathop{\famzero exp}\nolimits}

\def\lim{\mathop{\famzero lim}}

\def\log{\mathop{\famzero log}\nolimits}
\def\max{\mathop{\famzero max}}

\def\textfont@#1#2{\def#1{\relax\ifmmode
    \errmessage{Use \string#1\space only in text}\else#2\fi}}
\textfont@\rm\tenrm
\textfont@\it\tenit
\textfont@\sl\tensl
\textfont@\bf\tenbf
\textfont@\smc\tensmc
\let\ic@=\/
\def\/{\unskip\ic@}
\def\textfonti{\the\textfont1 }
\def\t#1#2{{\edef\next{\the\font}\textfonti\accent"7F \next#1#2}}
\let\B=\=
\let\D=\.
\def~{\unskip\nobreak\ \ignorespaces}
{\catcode`\@=\active
\gdef\@{\char'100 }}
\atdef@-{\leavevmode\futurelet\next\athyph@}
\def\athyph@{\ifx\next-\let\next=\athyph@@
  \else\let\next=\athyph@@@\fi\next}
\def\athyph@@@{\hbox{-}}
\def\athyph@@#1{\futurelet\next\athyph@@@@}
\def\athyph@@@@{\if\next-\def\next##1{\hbox{---}}\else
    \def\next{\hbox{--}}\fi\next}
\def\.{.\spacefactor=\@m}
\atdef@.{\null.}
\atdef@,{\null,}
\atdef@;{\null;}
\atdef@:{\null:}
\atdef@?{\null?}
\atdef@!{\null!}
\def\srdr@{\thinspace}
\def\drsr@{\kern.02778em}
\def\sldl@{\kern.02778em}
\def\dlsl@{\thinspace}
\atdef@"{\unskip\futurelet\next\atqq@}
\def\atqq@{\ifx\next\Space@\def\next. {\atqq@@}\else
         \def\next.{\atqq@@}\fi\next.}
\def\atqq@@{\futurelet\next\atqq@@@}
\def\atqq@@@{\ifx\next`\def\next`{\atqql@}\else\def\next'{\atqqr@}\fi\next}
\def\atqql@{\futurelet\next\atqql@@}
\def\atqql@@{\ifx\next`\def\next`{\sldl@``}\else\def\next{\dlsl@`}\fi\next}
\def\atqqr@{\futurelet\next\atqqr@@}
\def\atqqr@@{\ifx\next'\def\next'{\srdr@''}\else\def\next{\drsr@'}\fi\next}

\def\textfontii{\the\textfont2 }
\def\{{\relax\ifmmode\lbrace\else
    {\textfontii f}\spacefactor=\@m\fi}
\def\}{\relax\ifmmode\rbrace\else
    \let\@sf=\empty\ifhmode\edef\@sf{\spacefactor=\the\spacefactor}\fi
      {\textfontii g}\@sf\relax\fi}
\def\nonhmodeerr@#1{\errmessage
     {\string#1\space allowed only within text}}
\def\linebreak{\relax\ifhmode\unskip\break\else
    \nonhmodeerr@\linebreak\fi}
\def\allowlinebreak{\relax
   \ifhmode\allowbreak\else\nonhmodeerr@\allowlinebreak\fi}
\newskip\saveskip@
\def\nolinebreak{\relax\ifhmode\saveskip@=\lastskip\unskip
  \nobreak\ifdim\saveskip@>\z@\hskip\saveskip@\fi
   \else\nonhmodeerr@\nolinebreak\fi}
\def\newline{\relax\ifhmode\null\hfil\break
    \else\nonhmodeerr@\newline\fi}
\def\nonmathaerr@#1{\errmessage
     {\string#1\space is not allowed in display math mode}}
\def\nonmathberr@#1{\errmessage{\string#1\space is allowed only in math mode}}
\def\mathbreak{\relax\ifmmode\ifinner\break\else
   \nonmathaerr@\mathbreak\fi\else\nonmathberr@\mathbreak\fi}
\def\nomathbreak{\relax\ifmmode\ifinner\nobreak\else
    \nonmathaerr@\nomathbreak\fi\else\nonmathberr@\nomathbreak\fi}
\def\allowmathbreak{\relax\ifmmode\ifinner\allowbreak\else
     \nonmathaerr@\allowmathbreak\fi\else\nonmathberr@\allowmathbreak\fi}
\def\pagebreak{\relax\ifmmode
   \ifinner\errmessage{\string\pagebreak\space
     not allowed in non-display math mode}\else\postdisplaypenalty-\@M\fi
   \else\ifvmode\penalty-\@M\else\edef\spacefactor@
       {\spacefactor=\the\spacefactor}\vadjust{\penalty-\@M}\spacefactor@
        \relax\fi\fi}
\def\nopagebreak{\relax\ifmmode
     \ifinner\errmessage{\string\nopagebreak\space
    not allowed in non-display math mode}\else\postdisplaypenalty\@M\fi
    \else\ifvmode\nobreak\else\edef\spacefactor@
        {\spacefactor=\the\spacefactor}\vadjust{\penalty\@M}\spacefactor@
         \relax\fi\fi}
\def\newpage{\relax\ifvmode\vfill\penalty-\@M\else\nonvmodeerr@\newpage\fi}
\def\nonvmodeerr@#1{\errmessage
    {\string#1\space is allowed only between paragraphs}}
\def\smallpagebreak{\relax\ifvmode\smallbreak
      \else\nonvmodeerr@\smallpagebreak\fi}
\def\medpagebreak{\relax\ifvmode\medbreak
       \else\nonvmodeerr@\medpagebreak\fi}
\def\bigpagebreak{\relax\ifvmode\bigbreak
      \else\nonvmodeerr@\bigpagebreak\fi}
\newdimen\captionwidth@
\captionwidth@=\hsize
\advance\captionwidth@ by -1.5in
\def\caption#1{}
\def\topspace#1{\gdef\thespace@{#1}\ifvmode\def\next
    {\futurelet\next\topspace@}\else\def\next{\nonvmodeerr@\topspace}\fi\next}
\def\topspace@{\ifx\next\Space@\def\next. {\futurelet\next\topspace@@}\else
     \def\next.{\futurelet\next\topspace@@}\fi\next.}
\def\topspace@@{\ifx\next\caption\let\next\topspace@@@\else
    \let\next\topspace@@@@\fi\next}
 \def\topspace@@@@{\topinsert\vbox to
       \thespace@{}\endinsert}
\def\topspace@@@\caption#1{\topinsert\vbox to
    \thespace@{}\nobreak
      \smallskip
    \setbox\z@=\hbox{\noindent\ignorespaces#1\unskip}%
   \ifdim\wd\z@>\captionwidth@
   \centerline{\vbox{\hsize=\captionwidth@\noindent\ignorespaces#1\unskip}}%
   \else\centerline{\box\z@}\fi\endinsert}
\def\midspace#1{\gdef\thespace@{#1}\ifvmode\def\next
    {\futurelet\next\midspace@}\else\def\next{\nonvmodeerr@\midspace}\fi\next}
\def\midspace@{\ifx\next\Space@\def\next. {\futurelet\next\midspace@@}\else
     \def\next.{\futurelet\next\midspace@@}\fi\next.}
\def\midspace@@{\ifx\next\caption\let\next\midspace@@@\else
    \let\next\midspace@@@@\fi\next}
 \def\midspace@@@@{\midinsert\vbox to
       \thespace@{}\endinsert}
\def\midspace@@@\caption#1{\midinsert\vbox to
    \thespace@{}\nobreak
      \smallskip
      \setbox\z@=\hbox{\noindent\ignorespaces#1\unskip}%
      \ifdim\wd\z@>\captionwidth@
    \centerline{\vbox{\hsize=\captionwidth@\noindent\ignorespaces#1\unskip}}%
    \else\centerline{\box\z@}\fi\endinsert}
\mathchardef\prime@="0230
\def\prime{{{}\prime@{}}}
\def\prim@s{\prime@\futurelet\next\pr@m@s}

\def\,{\relax\ifmmode\mskip\thinmuskip\else\thinspace\fi}
\def\!{\relax\ifmmode\mskip-\thinmuskip\else\negthinspace\fi}
\def\frac#1#2{{#1\over#2}}

\def\:{\nobreak\hskip.1111em{:}\hskip.3333em plus .0555em\relax}
\def\intic@{\mathchoice{\hskip5\p@}{\hskip4\p@}{\hskip4\p@}{\hskip4\p@}}
\def\negintic@
 {\mathchoice{\hskip-5\p@}{\hskip-4\p@}{\hskip-4\p@}{\hskip-4\p@}}
\def\intkern@{\mathchoice{\!\!\!}{\!\!}{\!\!}{\!\!}}
\def\intdots@{\mathchoice{\cdots}{{\cdotp}\mkern1.5mu
    {\cdotp}\mkern1.5mu{\cdotp}}{{\cdotp}\mkern1mu{\cdotp}\mkern1mu
      {\cdotp}}{{\cdotp}\mkern1mu{\cdotp}\mkern1mu{\cdotp}}}
\newcount\intno@
\def\iint{\intno@=\tw@\futurelet\next\ints@}
\def\iiint{\intno@=\thr@@\futurelet\next\ints@}
\def\iiiint{\intno@=4 \futurelet\next\ints@}
\def\idotsint{\intno@=\z@\futurelet\next\ints@}
\def\ints@{\findlimits@\ints@@}
\newif\iflimtoken@
\newif\iflimits@
\def\findlimits@{\limtoken@false\limits@false\ifx\next\limits
 \limtoken@true\limits@true\else\ifx\next\nolimits\limtoken@true\limits@false
    \fi\fi}
\def\multintlimits@{\intop\ifnum\intno@=\z@\intdots@
  \else\intkern@\fi
    \ifnum\intno@>\tw@\intop\intkern@\fi
     \ifnum\intno@>\thr@@\intop\intkern@\fi\intop}
\def\multint@{\int\ifnum\intno@=\z@\intdots@\else\intkern@\fi
   \ifnum\intno@>\tw@\int\intkern@\fi
    \ifnum\intno@>\thr@@\int\intkern@\fi\int}
\def\ints@@{\iflimtoken@\def\ints@@@{\iflimits@
   \negintic@\mathop{\intic@\multintlimits@}\limits\else
    \multint@\nolimits\fi\eat@}\else
     \def\ints@@@{\multint@\nolimits}\fi\ints@@@}
\def\Sb{_\bgroup\vspace@
        \baselineskip=\fontdimen10 \scriptfont\tw@
        \advance\baselineskip by \fontdimen12 \scriptfont\tw@
        \lineskip=\thr@@\fontdimen8 \scriptfont\thr@@
        \lineskiplimit=\thr@@\fontdimen8 \scriptfont\thr@@
        \Let@\vbox\bgroup\halign\bgroup \hfil$\scriptstyle
            {##}$\hfil\cr}
\def\endSb{\crcr\egroup\egroup\egroup}
\def\Sp{^\bgroup\vspace@
        \baselineskip=\fontdimen10 \scriptfont\tw@
        \advance\baselineskip by \fontdimen12 \scriptfont\tw@
        \lineskip=\thr@@\fontdimen8 \scriptfont\thr@@
        \lineskiplimit=\thr@@\fontdimen8 \scriptfont\thr@@
        \Let@\vbox\bgroup\halign\bgroup \hfil$\scriptstyle
            {##}$\hfil\cr}
\def\endSp{\crcr\egroup\egroup\egroup}
\def\Let@{\relax\iffalse{\fi\let\\=\cr\iffalse}\fi}
\def\vspace@{\def\vspace##1{\noalign{\vskip##1 }}}
\def\aligned{\,\vcenter\bgroup\vspace@\Let@\openup\jot\m@th\ialign
  \bgroup \strut\hfil$\displaystyle{##}$&$\displaystyle{{}##}$\hfil\crcr}
\def\endaligned{\crcr\egroup\egroup}
\def\matrix{\,\vcenter\bgroup\Let@\vspace@
    \normalbaselines
  \m@th\ialign\bgroup\hfil$##$\hfil&&\quad\hfil$##$\hfil\crcr
    \mathstrut\crcr\noalign{\kern-\baselineskip}}
\def\endmatrix{\crcr\mathstrut\crcr\noalign{\kern-\baselineskip}\egroup
                \egroup\,}
\newtoks\hashtoks@
\hashtoks@={#}
\def\format{\crcr\egroup\iffalse{\fi\ifnum`}=0 \fi\format@}
\def\format@#1\\{\def\preamble@{#1}%
  \def\c{\hfil$\the\hashtoks@$\hfil}%
  \def\r{\hfil$\the\hashtoks@$}%
  \def\l{$\the\hashtoks@$\hfil}%
  \setbox\z@=\hbox{\xdef\Preamble@{\preamble@}}\ifnum`{=0 \fi\iffalse}\fi
   \ialign\bgroup\span\Preamble@\crcr}

\def\cases{\left\{\,\vcenter\bgroup\vspace@
     \normalbaselines\openup\jot\m@th
       \Let@\ialign\bgroup$##$\hfil&\quad$##$\hfil\crcr
      \mathstrut\crcr\noalign{\kern-\baselineskip}}

\newif\iftagsleft@
\tagsleft@true
\def\TagsOnRight{\global\tagsleft@false}
\def\tag#1$${\iftagsleft@\leqno\else\eqno\fi
 \hbox{\def\pagebreak{\global\postdisplaypenalty-\@M}%
 \def\nopagebreak{\global\postdisplaypenalty\@M}\rm(#1\unskip)}%
  $$\postdisplaypenalty\z@\ignorespaces}
\interdisplaylinepenalty=\@M
\def\allowdisplaybreak@{\def\allowdisplaybreak{\noalign{\allowbreak}}}
\def\displaybreak@{\def\displaybreak{\noalign{\break}}}
\def\align#1\endalign{\def\tag{&}\vspace@\allowdisplaybreak@\displaybreak@
  \iftagsleft@\lalign@#1\endalign\else
   \ralign@#1\endalign\fi}
\def\ralign@#1\endalign{\displ@y\Let@\tabskip\centering\halign to\displaywidth
     {\hfil$\displaystyle{##}$\tabskip=\z@&$\displaystyle{{}##}$\hfil
       \tabskip=\centering&\llap{\hbox{(\rm##\unskip)}}\tabskip\z@\crcr
             #1\crcr}}
\def\lalign@
 #1\endalign{\displ@y\Let@\tabskip\centering\halign to \displaywidth
   {\hfil$\displaystyle{##}$\tabskip=\z@&$\displaystyle{{}##}$\hfil
   \tabskip=\centering&\kern-\displaywidth
        \rlap{\hbox{(\rm##\unskip)}}\tabskip=\displaywidth\crcr
               #1\crcr}}
\def\overrightarrow{\mathpalette\overrightarrow@}
\def\overrightarrow@#1#2{\vbox{\ialign{$##$\cr
    #1{-}\mkern-6mu\cleaders\hbox{$#1\mkern-2mu{-}\mkern-2mu$}\hfill
     \mkern-6mu{\to}\cr
     \noalign{\kern -1\p@\nointerlineskip}
     \hfil#1#2\hfil\cr}}}
\def\overleftarrow{\mathpalette\overleftarrow@}
\def\overleftarrow@#1#2{\vbox{\ialign{$##$\cr
     #1{\leftarrow}\mkern-6mu\cleaders\hbox{$#1\mkern-2mu{-}\mkern-2mu$}\hfill
      \mkern-6mu{-}\cr
     \noalign{\kern -1\p@\nointerlineskip}
     \hfil#1#2\hfil\cr}}}
\def\overleftrightarrow{\mathpalette\overleftrightarrow@}
\def\overleftrightarrow@#1#2{\vbox{\ialign{$##$\cr
     #1{\leftarrow}\mkern-6mu\cleaders\hbox{$#1\mkern-2mu{-}\mkern-2mu$}\hfill
       \mkern-6mu{\to}\cr
    \noalign{\kern -1\p@\nointerlineskip}
      \hfil#1#2\hfil\cr}}}
\def\underrightarrow{\mathpalette\underrightarrow@}
\def\underrightarrow@#1#2{\vtop{\ialign{$##$\cr
    \hfil#1#2\hfil\cr
     \noalign{\kern -1\p@\nointerlineskip}
    #1{-}\mkern-6mu\cleaders\hbox{$#1\mkern-2mu{-}\mkern-2mu$}\hfill
     \mkern-6mu{\to}\cr}}}
\def\underleftarrow{\mathpalette\underleftarrow@}
\def\underleftarrow@#1#2{\vtop{\ialign{$##$\cr
     \hfil#1#2\hfil\cr
     \noalign{\kern -1\p@\nointerlineskip}
     #1{\leftarrow}\mkern-6mu\cleaders\hbox{$#1\mkern-2mu{-}\mkern-2mu$}\hfill
      \mkern-6mu{-}\cr}}}
\def\underleftrightarrow{\mathpalette\underleftrightarrow@}
\def\underleftrightarrow@#1#2{\vtop{\ialign{$##$\cr
      \hfil#1#2\hfil\cr
    \noalign{\kern -1\p@\nointerlineskip}
     #1{\leftarrow}\mkern-6mu\cleaders\hbox{$#1\mkern-2mu{-}\mkern-2mu$}\hfill
       \mkern-6mu{\to}\cr}}}
\def\sqrt#1{\radical"270370 {#1}}
\def\dots{\relax\ifmmode\let\next=\ldots\else\let\next=\tdots@\fi\next}
\def\tdots@{\unskip\ \tdots@@}
\def\tdots@@{\futurelet\next\tdots@@@}
\def\tdots@@@{$\mathinner{\ldotp\ldotp\ldotp}\,
   \ifx\next,$\else
   \ifx\next.\,$\else
   \ifx\next;\,$\else
   \ifx\next:\,$\else
   \ifx\next?\,$\else
   \ifx\next!\,$\else
   $ \fi\fi\fi\fi\fi\fi}
\def\text{\relax\ifmmode\let\next=\text@\else\let\next=\text@@\fi\next}
\def\text@@#1{\hbox{#1}}
\def\text@#1{\mathchoice
 {\hbox{\everymath{\displaystyle}\def\textfonti{\the\textfont1 }%
    \def\textfontii{\the\textfont2 }\textdef@@ T#1}}
 {\hbox{\everymath{\textstyle}\def\textfonti{\the\textfont1 }%
    \def\textfontii{\the\textfont2 }\textdef@@ T#1}}
 {\hbox{\everymath{\scriptstyle}\def\textfonti{\the\scriptfont1 }%
   \def\textfontii{\the\scriptfont2 }\textdef@@ S\rm#1}}
 {\hbox{\everymath{\scriptscriptstyle}\def\textfonti{\the\scriptscriptfont1 }%
   \def\textfontii{\the\scriptscriptfont2 }\textdef@@ s\rm#1}}}
\def\textdef@@#1{\textdef@#1\rm \textdef@#1\bf
   \textdef@#1\sl \textdef@#1\it}

\def\textdef@#1#2{\def\next{\csname\expandafter\eat@\string#2fam\endcsname}%
\if S#1\edef#2{\the\scriptfont\next\relax}%
 \else\if s#1\edef#2{\the\scriptscriptfont\next\relax}%
 \else\edef#2{\the\textfont\next\relax}\fi\fi}
\scriptfont\itfam=\tenit \scriptscriptfont\itfam=\tenit
\scriptfont\slfam=\tensl \scriptscriptfont\slfam=\tensl
\mathcode`\0="0030
\mathcode`\1="0031
\mathcode`\2="0032
\mathcode`\3="0033
\mathcode`\4="0034
\mathcode`\5="0035
\mathcode`\6="0036
\mathcode`\7="0037
\mathcode`\8="0038
\mathcode`\9="0039
\def\Cal{\relax\ifmmode\let\next=\Cal@\else
     \def\next{\errmessage{Use \string\Cal\space only in math mode}}\fi\next}
\def\Cal@#1{{\fam2 #1}}
\def\bold{\relax\ifmmode\let\next=\bold@\else
   \def\next{\errmessage{Use \string\bold\space only in math
      mode}}\fi\next}\def\bold@#1{{\fam\bffam #1}}
\mathchardef\Gamma="0000
\mathchardef\Delta="0001
\mathchardef\Theta="0002
\mathchardef\Lambda="0003
\mathchardef\Xi="0004
\mathchardef\Pi="0005
\mathchardef\Sigma="0006
\mathchardef\Upsilon="0007
\mathchardef\Phi="0008
\mathchardef\Psi="0009
\mathchardef\Omega="000A
\mathchardef\varGamma="0100
\mathchardef\varDelta="0101
\mathchardef\varTheta="0102
\mathchardef\varLambda="0103
\mathchardef\varXi="0104
\mathchardef\varPi="0105
\mathchardef\varSigma="0106
\mathchardef\varUpsilon="0107
\mathchardef\varPhi="0108
\mathchardef\varPsi="0109
\mathchardef\varOmega="010A
\font\dummyft@=dummy
\fontdimen1 \dummyft@=\z@
\fontdimen2 \dummyft@=\z@
\fontdimen3 \dummyft@=\z@
\fontdimen4 \dummyft@=\z@
\fontdimen5 \dummyft@=\z@
\fontdimen6 \dummyft@=\z@
\fontdimen7 \dummyft@=\z@
\fontdimen8 \dummyft@=\z@
\fontdimen9 \dummyft@=\z@
\fontdimen10 \dummyft@=\z@
\fontdimen11 \dummyft@=\z@
\fontdimen12 \dummyft@=\z@
\fontdimen13 \dummyft@=\z@
\fontdimen14 \dummyft@=\z@
\fontdimen15 \dummyft@=\z@
\fontdimen16 \dummyft@=\z@
\fontdimen17 \dummyft@=\z@
\fontdimen18 \dummyft@=\z@
\fontdimen19 \dummyft@=\z@
\fontdimen20 \dummyft@=\z@
\fontdimen21 \dummyft@=\z@
\fontdimen22 \dummyft@=\z@
\def\fontlist@{\\{\tenrm}\\{\sevenrm}\\{\fiverm}\\{\teni}\\{\seveni}%
 \\{\fivei}\\{\tensy}\\{\sevensy}\\{\fivesy}\\{\tenex}\\{\tenbf}\\{\sevenbf}%
 \\{\fivebf}\\{\tensl}\\{\tenit}\\{\tensmc}}
\def\dodummy@{{\def\\##1{\global\let##1=\dummyft@}\fontlist@}}
\newif\ifsyntax@
\newcount\countxviii@
\def\newtoks@{\alloc@5\toks\toksdef\@cclvi}
\def\nopages@{\output={\setbox\z@=\box\@cclv \deadcycles=\z@}\newtoks@\output}
\def\syntax{\syntax@true\dodummy@\countxviii@=\count18
\loop \ifnum\countxviii@ > \z@ \textfont\countxviii@=\dummyft@
   \scriptfont\countxviii@=\dummyft@ \scriptscriptfont\countxviii@=\dummyft@
     \advance\countxviii@ by-\@ne\repeat
\dummyft@\tracinglostchars=\z@
  \nopages@\frenchspacing\hbadness=\@M}
\def\magstep#1{\ifcase#1 1000\or
 1200\or 1440\or 1728\or 2074\or 2488\or
 \errmessage{\string\magstep\space only works up to 5}\fi\relax}
{\lccode`\2=`\p \lccode`\3=`\t
 \lowercase{\gdef\tru@#123{#1truept}}}

\def\scaletype#1{\mag=#1\relax
 \hsize=\expandafter\tru@\the\hsize
 \vsize=\expandafter\tru@\the\vsize
 \dimen\footins=\expandafter\tru@\the\dimen\footins}

\def\scalefont#1#2\andcallit#3{\edef\font@{\the\font}#1\font#3=
  \fontname\font\space scaled #2\relax\font@}
\def\Mag@#1#2{\ifdim#1<1pt\multiply#1 #2\relax\divide#1 1000 \else
  \ifdim#1<10pt\divide#1 10 \multiply#1 #2\relax\divide#1 100\else
  \divide#1 100 \multiply#1 #2\relax\divide#1 10 \fi\fi}
\def\scalelinespacing#1{\Mag@\baselineskip{#1}\Mag@\lineskip{#1}%
  \Mag@\lineskiplimit{#1}}
\def\wlog#1{\immediate\write-1{#1}}
\catcode`\@=\active

\font\tenbf=cmbx10
\font\tenrm=cmr10
\font\tenit=cmti10

\font\ninerm=cmr9

\font\eightbf=cmbx8
\font\eightrm=cmr8
\font\eightit=cmti8
\font\sevenrm=cmr7
\TagsOnRight
\hsize=5.0in
\vsize=7.4in
\parindent=15pt
\overfullrule=0pt
\nopagenumbers
\baselineskip=10pt

\def\sectiontitle#1\par{\bigskip\leftline{\bf #1}\nobreak\vglue 5pt}
\outer\def\state #1. #2\par{\medbreak
 \noindent{\tenbf#1. \enspace}{\sl#2}\par
\ifdim\lastskip
<6pt\removelastskip\penalty55\medskip\fi}

%
%
%
%
%
%
%
%

\def\qed{\hbox{${\vcenter{\vbox{
    \hrule height 0.4pt\hbox{\vrule width 0.4pt height 6pt
    \kern5pt\vrule width 0.4pt}\hrule height 0.4pt}}}$}}
\def\proof#1\par{\medbreak\noindent{\it Proof.\/}\quad{#1}\hfill\llap{\qed}\par
    \ifdim\lastskip<\medskipamount\removelastskip\penalty55\medskip\fi}

\def\pproof#1\par%
{\medbreak\noindent{\it Proof.\/}\quad{#1}\par
    \ifdim\lastskip<\medskipamount\removelastskip\penalty55\medskip\fi}

\def\for{\  \hbox{ for } \ }
\def\if{ \ \hbox{ if } \ }

\def\where{\  \hbox{ where } \ }
\def\and{\  \hbox{ and } \ }
\def\or{\  \hbox{ or } \ }

\def\equal{\ \displaystyle{\mathop  = ^{\hbox{\rm def}}}\ }

\def\la{\lambda}

\def\om{ {\omega}}

\def\th{\theta}
\def\al{\alpha}
\def\be{\beta}

\def\ep{\epsilon}

\def\de{\delta}
\def\De{\Delta}
\def\ka{\eta}
\def\si{\sigma}

\def\ze{\zeta}

\def\pa{\partial}

\def\vph{\varphi}

\def\vep{\varepsilon}
\def\vpi{{\varpi}}
\def\vth{{\vartheta}}
\def\vsi{{\varsigma}}
\def\vrh{{\varrho}}

\def\tal{\tilde{\alpha}}
\def\tbe{\tilde{\beta}}
\def\tga{\tilde{\gamma}}
\def\tvs{\tilde{{\varsigma}}}
\def\tu{\tilde{u}}
\def\tw{\tilde w}

\def\tB{\tilde B}
\def\tv{\tilde v}
\def\tz{\tilde z}
\def\tb{\tilde b}
\def\tB{\tilde B}

\def\trh{\tilde {\rho}}

\def\tr{\tilde r}
\def\tP{\tilde P}

\def\hH{\hat{H}}
\def\hh{\hat{h}}

\def\hu{\hat{u}}
\def\hs{\hat{s}}
\def\hv{\hat{v}}
\def\hb{\hat{b}}

\def\CC{{\hbox{\eightbf C}}}
\def\C{\hbox{\bf C}}

\def\R{\hbox{\bf R}}
\def\bs{\hbox{\bf S}}
\def\Z{\hbox{\bf Z}}

\def\D{{\Cal D}}
\def\B{{\Cal B}}

\def\f{\Cal{F}}
\def\t{\Cal{T}}
\def\r{\Cal{R}}
\def\l{\Cal{L}}
\def\m{\Cal{M}}
\def\k{\Cal{K}}
\def\n{\Cal{N}}

\def\p{\Cal{P}}

\def\c{\Cal{C}}

\def\v{\Cal{V}}

\def\u{\Cal{U}}

\def\End{\mathop{\hbox{\rm End}\,}_{\hbox{\eightbf C}}\,}
\def\Hom{\mathop{\hbox{\rm Hom}\,}_{\hbox{\eightbf C}}\,}
\def\dim{\mathop{\hbox{\rm dim}\,}_{\hbox{\eightbf C}}\,}
\def\Aut{\mathop{\hbox{\rm Aut}\,}_{\hbox{\eightbf C}}\,}
\def\ct{\hbox{\rm ct}}
\def\Red{\hbox{\rm Red}}

\def\0{\bold{0}}

\def\gl{\goth{gl}}

\font\germ=eufb10 
\def\goth#1{\hbox{\germ #1}}

\def\H{\goth{H}}  

\font\smm=msbm10 at 12pt 
\def\symbol#1{\hbox{\smm #1}}
\def\lsmash{{\symbol n}}

%
%
%
%
%

%
%
%

\line{\eightrm
Submitted to Communs Math. Phys.\hfil}
\phantom{
\line
{\eightrm $\copyright$\, World Scientific Publishing Company \hfil}
}

\headline={\ifnum\pageno=1\hfil\else%
{\ifodd\pageno\rightheadline \else \leftheadline\fi}\fi}
\def\rightheadline{\hfil\eightit Elliptic
 Quantum Many-Body Problem \quad\eightrm\folio}
\def\leftheadline{\eightrm\folio\quad
\eightit Ivan Cherednik\hfil}
\voffset=2\baselineskip
\vglue 5pc
\baselineskip=13pt
\centerline{\tenbf ELLIPTIC QUANTUM MANY-BODY PROBLEM}
\centerline{\tenbf AND DOUBLE AFFINE KNIZHNIK-ZAMOLODCHIKOV EQUATION}
\vglue 24pt
\centerline{\eightrm IVAN CHEREDNIK\footnote"$^*$"
{\eightit \baselineskip=10pt
Partially supported by NSF grant number DMS--9301114 and UNC Research
Counsel grant}}

\baselineskip=12pt
\centerline{\eightrm
 Dept. of Math.,
UNC at Chapel Hill}
\centerline{\eightit Chapel Hill, N.C. 27599-3250, USA }

\vglue 20pt
\centerline{\eightrm ABSTRACT}
{\rightskip=1.5pc
\leftskip=1.5pc
\eightrm\parindent=1pc
The elliptic-matrix quantum Olshanetsky-Perelomov problem is
introduced for  arbitrary root systems by means of an elliptic
generalization of the Dunkl operators.
Its equivalence with  the double affine
generalization of the Knizhnik-Zamolodchikov
equation (in the induced representations) is established.

\vglue 3pt
$$
\vbox{\align
\hbox{Section 0.}&\quad \hbox{Introduction}\cr
\hbox{Section 1.}&\quad \hbox{Double Hecke algebras}\cr
\hbox{Section 2.}&\quad \hbox{Affine r-matrices}\cr
\hbox{Section 3.}&\quad \hbox{Dunkl operators and KZ}\cr
\hbox{Section 4.}&\quad \hbox{Examples}
\endalign}
$$
\vglue 3pt}
\baselineskip=13pt

\sectiontitle
0. Introduction

We generalize the affine
the Knizhnik-Zamolodchikov equation from [Ch1,2,3] replacing the
corresponding root systems by their affine counterparts.
To explain the construction
in the case of the root system of $\gl_n$, let us first introduce
the {\tenbf  affine Weyl group} $\bs_n^a$. It is the semi-direct
product of the symmetric group $\bs_n$ and the lattice
$A \ =\ \oplus _{i=1}^{n-1} \Z \ep_{i i+1} $,
where the first acts on the second
permuting $\{\ep_i, \ \ep_{ij}\ =\ \ep_i-\ep_j\}$ naturally.
This group is generated by the adjacent transpositions
$$s_i = (i i+1), \ 1\le i< n, \and
s_0 = s_{n1}^{[1]}, \hbox{\ where\ \ }
s_{ij}^{[k]}\ =\  (ij)(k\ep_{ij})\in \bs_n^a.
$$
Setting
$$
\align
&s_{ij}^{[k]}(b) = b- (\ep_{ij},b) (\ep_{ij}+kc),\
s_{ij}^{[k]}(c) = c,\quad b\in B= \oplus _{i=1}^{n} \Z \ep_i,
\tag0.1
\endalign
$$
we obtain an action of $\bs_n^a$ in $\tB\ =
\ B \oplus \Z c.\ $
In particular,
$$
\align
&s_0(b) = b+(b,\ep_{1n})(c - \ep_{1n}),\
a(\ep_i) = \ep_i - (a,\ep_i)c,\quad 1\le i\le n.
\endalign
$$
Put
$$
x_{  \tb}\ =\  k x_c+\sum_{i=1}^n k_i x_i,\
z_{  \tb}\ =\  k \xi+\sum_{i=1}^n k_i z_i
 \for  \tb\ = k c+ \sum_{i=1}^n k_i\ep_i.
$$

The {\tenbf double affine degenerate (graded) algebra} $\H'$
is generated by the group algebra $\C [\bs_n^a]$,
 pairwise commutative elements $\{x_b,\ b\in B\}$, and central $x_c$,
satisfying the relations (depending on
$\ka\in \C$):
$$
\align
&s_i x_b\ =\ x_{s_i(b)}s_i + \ka (\ep_{ii+1},b),\
 1\le i< n,\quad
  s_0 x_b\ = \ x_{s_0(b)}s_0 + \ka (\ep_{n1},b).
\tag0.2
\endalign
$$
This  algebra is an affine version of that considered
by Drinfeld and Lusztig and a degeneration of the double affine Hecke
algebra  from [Ch8] (for $\gl_n$).

Let us fix $\mu\in \C$ and set
$$ \hbox{\rm ct}_{ij}^{[k]}\ =\ \hbox{\rm ct}(z_{ij} +k\xi) \for
\hbox{\rm ct}(t)\ =\ (\exp (t)-1)^{-1},\ z_{ij}=z_i-z_j.$$
We introduce the differential operators of the first order:$$
\eqalignno{
\D_{\ep_i}=\D _i &= \pa/\pa z_i -\ka\sum _{n\ge j>i}\hbox{\rm
ct}_{ij}^{[0]}(s_{ij}^{[0]}-\mu)
+ \ka\sum _{1\le j<i}\hbox{\rm ct}_{ji}^{[0]}(s_{ji}^{[0]}-\mu)\cr
&-\ka\sum _{j\neq i}\sum_{k>0}
\Bigl( \hbox{\rm ct}_{ij}^{[k]}(s_{ij}^{[k]}-\mu)
 - \hbox{\rm ct}_{ji}^{[k]}(s_{ji}^{[k]}-\mu)\Bigr)\ +\
\mu \ka (n/2-i+1),\cr
\D_c\ &=\ \pa/\pa \ze +\ka\mu n, \quad 1\le i,j\le n.
&(0.3)
}
$$
We consider the sums formally as infinite linear combinations
of the elements  $\tw\in \bs_n^a$ with
the coefficients depending on
 $\{z,\xi\}$ and one
more complex variable $\ze$.
Assuming that $\Re(\xi)>0$, we can introduce a norm in this
space to
make all  series convergent.

The family of operators
$\{ \D'_i=\D_i-x_i,\ \D'_c=\D_c-x_c\}$ is commutative and
$\bs_n^a$-invariant
with respect to the following simultaneous
action of this group on the coefficients
(that are from $\H'$) and the arguments $\{ z_{\tb},\ze\}$:
$$
\align
&\tw(\hh)=\tw \hh \tw^{-1}, \ \hh\in\H',\
\tw(z_{ \tb})=z_{\{\tw( \tb)\}},\
\tb\in \tB,\cr
& s_i(\ze)\ =\ \ze \for 1\le i< n,\ s_0(\ze)\ =\ \ze-\xi+z_{1n}.
\tag0.4
\endalign
$$
The invariance means that $\tw (\D'_{\tu}) = \D'_{\tw (\tu)}$, where
$\D'_{\al \tu+\be \tv}\ = \ \al \D'_{ \tu}+\be \D'_{\tv}$
for  $\al,\be\in \Z,$
 $\tu,\tv\in  \tB,\and \tw\in \bs_n^a.$
Actually this family is invariant even with respect to the action of
the bigger
group generated by $W$ and $B$ (instead of $A$). It leads to
a natural extension of the above $\H'$. The precise choice of constants
in (0.3) is necessary to ensure the $B$-invariance.

The  {\tenbf double affine KZ}
is the system $\{\D'_{u}\Phi=0, \ u\in  B\}$
for a function $\Phi(z)$ with the values in $\H'$ or its representations.
Here $\xi$ is considered as a parameter.

Let us factorize $\H'$  by the ideal $(x_c)$.
The symmetric polynomials in $x_1,\ldots,x_n$
constitute the center of the resulting algebra   $\H'_0$.
Given a fixed central character and a
 finite dimensional $\C[\bs_n^a]$-modules $V$,
the corresponding induced (universal) $\H'_0$-module
is finite dimensional as well. When considered
in  this representation,
the  series in (0.3) become
convergent (at least for rather big $\Re(\xi)$)
and turn into  functions of elliptic type.
The corresponding
double KZ is equivalent to a $V$-valued
version of the
{\tenbf
elliptic quantum many-body problem} from [OP]
(which also generalizes the matrix QMB
 from [Ch5]).

To  introduce the latter
 let us first consider the same
formulas (0.3) assuming
now that $s_{ij}^{[k]}$ act on the arguments $\{z_{b},\ze\}$
as in (0.4). We will write $\si(\tw)$ and $\si_{ij}^{[k]}$
instead of $\tw$ and $s_{ij}^{[k]}$ to emphasize  this.
  The corresponding {\tenbf elliptic
Dunkl operators} (which are scalar but not pure differential anymore)
 will be denoted by $\{ \De_i, \De_c,\De_{\tb}\}$. The map
$$\tw\to\si(\tw),\ \ x_{\tb}\to \De_{\tb},\ \ \tw\in W^a,\
 \tb\in \tB,
$$
 gives a homomorphism
from the algebra $\H'$
into the algebra of operators acting on the space of (scalar)
 functions of $\{z,\ze\}$.
Imposing the relation $\De_c=0$  we obtain
 an embedding of $\H'_0$.

Second, given an  arbitrary symmetric polynomial
$p=p(x_1,\ldots,x_n)$,  we use (0.4) to represent
$$p(\De_1,\ldots,\De_n)\ =\ \sum_{\tw\in W^a}D_{\tw}\si(\tw),
\where D _{\tw} \hbox{\ are \ differential.}
$$
Then we replace  every $\si(\tw)$ by the image of $\tw^{-1}$
 in $\Aut V$ setting  $\ \pa/\pa\ze=-\ka\mu n\ $ afterwards.
The resulting operators
$\{L_p\}$ are $\bs_n^a$-invariant
and pairwise commutative. If $V$ is one-dimensional,
they coincide with the OP operators for $\mu=0$
and  are conjugated to them (by proper
remarkable scalar functions) when $\mu=\pm 1$ (with $V$ of
the same "sign").

 The element $p_2=\sum_{i=1}^n x_i^2$
 leads up to a constant
to the {\tenbf Schr\"odinger operator}
$$
\align
&H\ =\ \sum^n_{i=1}\pa^2/\pa z^2_i+
\hbox{const}\sum_{i<j}\wp(z_i-z_j)
\tag 0.5
\endalign
$$
in terms of the Weierstrass elliptic function with
the periods $\{(2\pi \imath), \xi\}$.

In this paper we consider arbitrary root systems and
any initial representations $V$ of  the corresponding
affine Weil groups.  We note that the
 commutative families of scalar $H$-operators for $A, B, D$ types
(with  certain  uniqueness theorems) were obtained recently by
direct methods (due to Heckman-Opdam)   in   [OOS].

It is worth mentioning that for  $\mu =1$
( and  special $\ka$)
the operators $L_p$ are expected to be
the radial parts of
Laplace operators for Kac-Moody symmetric spaces
at the critical level $c+n=0$.  The latter condition
gives the existence of the "big" center of
the corresponding universal enveloping algebra
(which is necessary to start the  Harish-Chandra,
Helgason theory of radial part). It is directly
connected with the  substitution
$\pa/\pa\ze=-\ka\mu n$.

Something can be done when $\ \pa/\pa\ze= \ka\mu m\ $ for arbitrary $m$.
 Let us introduce one more operator
$$
\eqalignno{
&\De_d\ =\ \pa /\pa{\xi}- \ka\sum _{i< j}\sum_{k>0} k\Bigl(
 \hbox{\rm ct}_{ij}^{[k]}(\si_{ij}^{[k]}-\mu)
 + \hbox{\rm ct}_{ji}^{[k]}(\si_{ji}^{[k]}-\mu)\Bigr).
&(0.6)
}
$$
It does not commute with $\{\De_i\}$, but
the operators $\De_{\hb}= \De_{b}+k\De_c+l\De_d $ for
$\hb= b +kc+ l d $ still satisfy the cross-relations:
$$
\align
&\si_i \De_{\hb}\ =\ \De_{s_i(\hb)}\si_i + \ka (\ep_{ii+1},\tb),\
 0\le i< n,\ \ep_{01}= c-\ep_{1n},\cr
&  s_i(c) = c,\
s_j(d) = d \for 1\le j< n,\
s_0(d) = d-\ep_{01},
 \tag0.7
\endalign
$$
where the form $(\ ,\  )$ is extended to $\R^{n+2}$
in the following way:
$$(c,c)=(c,\ep_i)=0=(d,\ep_i)=(d,d),\ 1\le i\le n,\ (c,d)=1.$$
It gives that  the operator $\ 2\De_{d}\De_c +\sum_{i=1}^n \De_{i}^2\ $
is $\bs_n$-invariant. Its reduction in the above sense
is also invariant
and is conjugated to $\hH = H+ 2\ka\mu (n +m)\pa/\pa\xi$
 in the setup
of (0.5).

When $\mu=1$ and $V$ is the corresponding
one-dimensional representation, $\hH$ was
introduced in [EK]. Presumably it is related to
the elliptic $r$-matrix KZ from
[Ch1] (with the additional equation from [E])
and   the so-called Bernard KZ equation
(see [FW,EK]). Certain
  remarkable properties of this  {\tenbf
parabolic operator} seem to be
important for the affine harmonic analysis
at arbitrary level.

In conclusion we would like to mention that all above constructions
have  difference counterparts and hopefully ensure
a basis for the elliptic
 Macdonald theory (see e.g. [M,O,Ch6]).

\par
This work was started at Weizmann Institute
(Israel). I thank A. Joseph and my colleagues at the institute
for the kind invitation
and hospitality. I am greatful to G.Felder for useful
discussions.

\sectiontitle
1. Double Hecke algebras

We follow [Ch3] (see also [Ch5,6,7]).
Reduced root systems only will be
discussed here. All the definitions and statements can be extended
to the general case. Minor changes in formulas
are necessary for divisible roots.

Given a Euclidean form ( , ) on $\R ^n$ and a root system
$R=\{ \alpha \} \subset \R ^n$ of type $A_n,B_n,\cdots ,G_2$,
let $s_\alpha $ be the orthogonal reflections in the hyperplanes
$(\alpha ,u)=0,\ u\in \R ^n$. Further, $\{ \alpha _1,\cdots \alpha _n\} $
are the simple roots relative to some fixed Weyl chamber, $R_+$
the set of all positive (written $\alpha >0)$ roots, $W$ the Weyl group
generated by $s_\alpha $ (or by
$s_i\displaystyle {\mathop =^{\hbox{\rm def}}}
s_{\alpha _i}, 1\le i\le n),\ \C [W]=\displaystyle {\oplus _w}\C w$
the group algebra of $W\ni w$.

We introduce $ a_i = \al_i^\vee,\where \al^\vee =2\al/(\al,\al)$,
the dual fundamental weights
$b_1,...,b_n$  satisfying the relations  $ (b_i,\al_j)=
\de_i^j$ for the
Kronecker delta, and the lattices
$$
\eqalignno{
& A=\oplus^n_{i=1}\Z a_i \subset B=\oplus^n_{i=1}\Z b_i,
}
$$

Let us fix a $W$-invariant set
$\ka=\{ \ka_\alpha \in \C , \alpha \in R\} $. The $W$-invariance
$({}^w\ka_\alpha =
\ka_{w(\alpha )}, w\in W)$ gives that $\ka_\alpha =\ka'$ or $\ka''$
respectively for the short and long roots ($R$ is supposed to
be reduced). We put $\ka_i=\ka_{\alpha _i}$ and define the $\ka$-
generalization of the $\rho$ and the Coxeter number $h$:
$$
\align
&2\rho _{\ka}=\sum_{\alpha >0}\ka_\alpha \alpha =\sum^n_{i=1}
\ka_i(\alpha _i,\alpha _i) b _i\in \R ^n, \cr
& h_{\ka}\ = \ \ka_{\th} + (\rho_{\ka}, \th)\for
\hbox{the maximal root\ } \th\in R_+.
\tag 1.1
\endalign
$$
We   will use
the same notations for  other $W$-invariant sets
 instead of $\ka$.

The following affine completion is common in the theory of
the Kac-Moody algebras (see e.g. [Ka], [Ch6]). Let us extend
the above pairing to $\R^{n+1}=\R^n\oplus \R c  $
setting $ (c,c)=0=(c,u)$.

The vectors (affine roots) $\ \tal=\al+kc $
for $\al \in R, k \in \Z,\ $
form the {\tenbf affine root system}
$R^a \supset R$.
We add  $\al_0 \equal c-\th $ to the  set of simple roots
and put $\ \ka_{\tal}= \ka_\al, \ \ka_0=\ka_{\th}=\ka''$.
The corresponding set $R^a_+$ of positive roots coincides
with $R_+\cup \{\al+kc, \al\in R, k > 0\}$. Let
$\tB\equal B\oplus \Z c.$
Given $\tal=\al+kc\in R^a,  \ a \in A,\  \tu=
 u+ \kappa  c \in \R^{n+1}$,
$$
\eqalignno{
&s_{\tal}(\tu)\ =\  \tu-(u,\al^{\vee})(\al + k c), \quad
a'(\tu)\ =\ \tu -(u,a)c.
&(1.2)
}
$$

The {\tenbf affine Weyl group} $W^a$ is generated by all $s_{\tal}$.
 One can take
the simple reflections $s_j=s_{\al_j}, 0 \le j \le n,$ as its
generators. This group is
the semi-direct product $W\lsmash A'$ of
its subgroups $W$ and $A'=\{a', a\in A\}$, where
$$
\eqalignno{
& a'=\ s_{\al}s_{\{\al+c\}}=\ s_{\{-\al+c\}}s_{\al}\for a=\al^{\vee},
 \al\in R .
}
$$

\state
Definition 1.1.
The {\tenbf degenerate (graded) double affine Hecke algebra} $\H '$ is
 algebraically generated by
 the group algebra $\C [W^a]$ and the pairwise commutative
$$
\align
&x_{\tu}\equal
\sum ^n_{i=1}(u,\alpha _i)x_i+ \kappa x_c\for
\tu=u+\kappa c \in \R^{n+1},\tag 1.3
\endalign
$$
satisfying  the following relations:
$$
\align
&s_ix_{\tu}-x_{\{s_i(\tu)\}}s_i\ =\ \ka_i(u,\al_i ),\quad 0\le i\le n.
\tag 1.4
\endalign
$$
The {\tenbf restricted  algebra} $\H'_0$ is
the factor-algebra $\H'/(x_c)$ (the quotient
by the central
ideal $(x_c)$).
\hfill \llap{\qed}

Without $i=0$ we arrive at the defining relations
$$
\align
&s_ix_i-(x_i-x_{a_i})s_i =  \ka_i,
\ s_ix_j=x_j s_i, \where
 1\le i\not= j\le n,\ a_i=\al_i^\vee,
\endalign
$$
of the graded affine Hecke algebra from [L] (see also [Ch3,5]).
 We mention that $\H'$ is a
degeneration of the double affine
Hecke algebras introduced in [Ch6,7].

Let $\C [x]\equal$ $ \C[x_1,\ldots,x_n,x_c]$ be
the algebra of polynomials in terms of $\{x_{\tu}\}$. We
denote the subalgebra
of $ W$-{\tenbf invariant polynomials} (with respect
to the  action of $W$ on $\{\tu\}$) by $\C[x]^W$. Later
 the same notations will be used
for other letters instead
of $x$.

\state
Theorem 1.2.
 An arbitrary element $\hh\in \H'$ can be uniquely represented
in the (left) form $\hh=\sum_{\tw\in W^a}f_{\tw}\tw$ and
the (right) form $\hh=\sum_{\tw\in W^a}\tw g_{\tw}$, where
$f_{\tw},g_{\tw}\in \C[x].$
 The center of
$\H '_0$ coincides with $\C[x]^W$.

{\it Proof.} The first statement  results from Theorem 2.3,
[Ch7] established in the non-degenerate case (see also [Ch6]).
Following [Ch3] one can check that the center of $\H'_0$
contains
$\C[x]^W$ and belongs to
$ \C [x]$ (that is  a  maximal commutative
subalgebra). The subalgebra
generated by $\C[W]$ and $ \C[x]$ is the
degenerate (graded) affine Hecke algebra
in the sense of [L,Ch3].
 Hence its center coincides with $\C[x]^W$
(due to Bernstein).
\par
\hfill \llap{\qed}

{\tenbf Induced representations. }
Let $V$ be a $\C [W^a]$-module, $V^o= \Hom (V,\C)$
 its dual with the natural action ($\tw(l(v))=l(\tw^{-1}v),
\ l\in  \Hom (V,\C)$),
$\tau\and \tau^o$ the corresponding homomorphisms from $\C [W^a]$
to $\ \End V\and \End V^o$. We  will use the diagonal action:
$$
\align
&\de(\tw)(v\otimes x_{\tu})=\tau^o (\tw)(v)\otimes \tw( x_{\tu}),\
\tw (x_{\tu})\ =\ x_{\{\tw(\tu)\}},\cr
&\for
v\otimes x_{\tu}\in \v\equal V^o\otimes _\CC \C [x],\
\tw\in W^a,\ \tu\in \R^n.
\tag1.5
\endalign
$$

The next proposition holds good for the entire
 $\H '$, but the latter has the trivial center $=\C x_c$
 (we need a "big" center to construct finite dimensional
representations).
Till the end of the section,   $x_c=0$ and  $x_{\tu}$ are identified
 with the corresponding
$x_u$.

\state
Proposition 1.3.
The universal (free) $\H'_0$-module generated by the $\C [W^a]$-module
$V^o$ is isomorphic to $\v$ with the natural action of $\C [x]$
by multiplications and the following action of $s_i$:
$$
\align
& \hs_i=\de(s_i)+ \ka_i x_{a_i}^{-1}(1-s _i), \  0\le i\le n,
\ a_0=c-\th^{\vee},
\tag 1.6
\endalign
$$
where $x_{a_i}^{-1}(1-s _i)(f)=
x_{a_i}^{-1}(f-s_i(f))$
for $f\in \v$ ($s_i$ acts only on $x$).

{\it Proof} follows [Ch3,5].
\hfill \llap{\qed}

We fix  a set $\la = \{\la_1,..., \la_n\}
\subset \C$ and consider the  quotient $\v_\la$ of $\v$
by the
(central) relations $p(x_1,..., x_n) =
p(\la_1,..., \la_n)$
for all $p\in \C [x]^W$.

Finally, we  introduce :
$$
\align
&V(\la)\equal(\v_{\la})^o = \hbox{\rm Hom}_\CC (\v_{\la},\C),
\ \hh(l(u))=l(\hh^o (u)), \
u\in \v_{\la}, l\in (\v_{\la})^o, \cr
&s^o_i=s_i, \  x^o_i=x_i, \
(\hh_1\hh_2)^o=\hh^o_2\hh^o_1,\ \ \hh_{1,2}\in \H'_0.\tag 1.7
\endalign
$$
The anti-involution $\hh\rightarrow \hh^o$
is well-defined because relations (1.4) are self-dual.

The above construction gives two canonical $W^a$-homomorphisms:
$$ id: \ V^o\rightarrow \v\rightarrow\v_{\la},\ \
tr: \ V(\la) \rightarrow V.
$$

\state
Proposition 1.4.
If a $\H '$-submodule $\u \subset V(\la)$ is non-zero
then its image $tr(\u)$ is non-zero too.

\proof It is clear, since $\v_{\la}$ is generated by $V^o$ as
an $\H '$-module.

 If $V$ is finite-dimensional
then $\dim V(\la)=|W| \dim V $, where $|W|$ is the
number of elements of $W$. The main examples will be for
one-dimensional representations of $W^a$ which are de\-scribed by
$W$-in\-va\-riant sets $\varepsilon$ $ \subset$ $\{ \pm 1\} $:
 $$
\align
&\tau _\varepsilon (s_i)=\varepsilon _i, \
\tau_\varepsilon (a')= 1,\ \ 0\le i\le n, \ a\in A.\tag 1.8
\endalign
$$
Let us denote the corresponding
$V,\ V(\la)$ by $\C_\vep,\ \C_\vep(\la)$ for the latter reference.

\sectiontitle
2. Affine $r$-matrices
\par
Following [Ch 1,3,5] we introduce abstract classical $r$-matrices
with the values in
 an arbitrary $\C$-algebra $\f$ and show how to extend
non-affine $r$-matrices to affine ones.
The notations are from Section 1. Let us denote
$\R\tal + \R\tbe \subset \R^n $
by $\R \langle \tal, \tbe \rangle $
 for $\tal,\tbe \in R^a $.

\state
Definition 2.1.
a) A set $r = \{ r_{\tal} \in \f, \tal \in
R^a_+ \} $ is an {\tenbf affine $r$-matrix} if
$$
\eqalignno{
& [r_{\tal}, r_{\tbe}] \ = \ 0,
 &(2.1)
}
$$
$$
\eqalignno{
 [r_{\tal}, r_{\tal+\tbe}]+ &[r_{\tal}, r_{\tbe}]+
[r_{\tal+\tbe},r_{\tbe}]\ =\ 0,
&(2.2) \cr
 [r_{\tal}, r_{\tal+\tbe}]  +
[r_{\tal},r_{\tbe}] +  &[r_{\tal+\tbe}, r_{\tal+2\tbe} ] + \cr
&[r_{\tal+\tbe},r_{\tbe}] +
 [r_{\tal+2\tbe}, r_{\tbe}] =  0,\ \
[r_{\tal}, r_{\tal+2\tbe} ] =0,
 &(2.3)
}
$$
$$
\eqalignno{
&[r_{\tal}, r_{3\tal+\tbe}]+
[r_{\tal},  r_{2\tal+\tbe}]+
[r_{\tal},  r_{3\tal+2\tbe}]+
[r_{\tal},  r_{\tal+\tbe}]+
[r_{\tal}, r_{\tbe}] +\cr
& [r_{3\tal+\tbe}, r_{2\tal+\tbe}]+
  [r_{2\tal+\tbe}, r_{3\tal+2\tbe}] +
  [r_{2\tal+\tbe}, r_{\tal+\tbe}] +
  [r_{3\tal+2\tbe}, r_{\tal+\tbe}]+
  [r_{\tal+\tbe}, r_{\tbe}] = 0,\cr
&[r_{3\tal+\tbe},  r_{3\tal+2\tbe}]+
 [r_{3\tal+\tbe}, r_{\tbe}] +
[r_{3\tal+2\tbe},  r_{\tbe}] = 0 =
 [r_{3\tal+\tbe}, r_{\tal+\tbe}] =
[r_{2\tal+\tbe},  r_{\tbe}],
&(2.4)
}
$$
under the assumption that $\tal, \tbe \in R^a_+ $ and
$$
\eqalignno{
&\R \langle \tal, \tbe \rangle \cap R^a = \{ \pm \tga \}, \tga \ \hbox{
 runs over all the indices}  &(2.5)
}
$$
in the corresponding identities.
\hfill\break
\indent
b) A {\tenbf closed $r$-matrix} (or a closure of the above $r$) is a
set $ \{ r_{\tal} \in \f, \tal \in R^a \} $ \  (extending $r$ and)
satisfying relations (2.1) - (2.4) for arbitrary (positive, negative)
$ \tal, \tbe \in R^a $ such that the corresponding condition  (2.5) is
fulfilled. If the indices are from $R_+$ (or $R$) we call $r$
{\tenbf non-affine}.
\par
\hfill \llap{\qed}

We note that (2.5) for identity (2.1) means that
$$
\eqalignno{
& (\tal, \tbe) = 0 \and \R \langle \tal, \tbe \rangle \cap R^a = \{ \pm \tal,
\pm \tbe \}.   &(2.6)
}
$$
It is equivalent to the existence of $\tw \in W^a$ such that
$ \tal = \tw(\al_i), \tbe = \tw(\al_j)$
for simple $ \al_i \ne \al_j \ (0\le i,j \le n) $ disconnected in the
affine Dynkin diagram of $R^a$.
In the most interesting
examples, (2.1) holds true for arbitrary orthogonal roots.

The corresponding assumptions for (2.2) - (2.4) give that $\tal, \tbe $
 are simple roots of a two-dimensional root subsystem in $R^a$
of type $A_2, B_2, G_2$. Here $\tal, \tbe$ stay for $\al_1 ,
\al_2$ in the notations from the figure of the systems of rank 2 from [B].
One can represent them as follows : $ \tal = \tw(\al_i), \tbe =
\tw(\al_j)$ for
a proper $\tw$ from $W^a$ and joined  (neighbouring )
$\al_i, \al_j$ .

 Given  an arbitrary $r$, we always have the following
closures (the standard one and the extension by zero):
$$
\eqalignno{
&r_{-\tal }\ =\ -r_{\tal},\quad
r_{-\tal }\ =\  0, \quad \tal \in R^a_+ .&(2.7)
}
$$
If there exists an action of $W^a\ni \tw$ on $\f$ such that
$$\tw(r_{\tal})\ =\ r_{\tw(\tal)} \for \tal,\ \tw(\tal)\in R^a_+,$$
then the extension of $r$ satisfying  these relations for
all $\tw$ is well-defined and closed (the invariant closure).

\state
Theorem 2.2.
 Let us assume  that  $r$ is a closed non-affine
$r$-matrix and  the group $A \ni a $ (see (1.2))
acts on the algebra
$\f \ni f $ (written $ f \to a(f) $) obeying the following
condition
$$
\eqalignno{
&a( r_{\al} ) = r_{\al} \quad\hbox{whenever}\quad (a,\al) = 0, a \in A,
\al \in R. &(2.8)
}
$$
Then the elements
$$
\eqalignno{
& r_{\tal} \equal a( r_{\al} ) \for a \hbox{\ such\ that\ \ }
 \tal = a'(\al) =
\al-(a,\al)c &(2.9)
}
$$
are well defined  (do not depend on the choice
of $a$ satisfying (2.9) for a given
 $\tal\in R^a$) and form a closed (affine)
$r$-matrix.

{\it Proof} is the same as that of Theorem 2.3 from [Ch4]
in the case of quantum $R$-matrices.
 \par
\hfill \llap{\qed}

\state
Theorem 2.3.
a)Given an affine $r$-matrix, let us suppose that the algebra $\f$ is
supplied (as a \C-linear space) with  a norm $||f||$ and
the following series are absolutely convergent:
$$
\eqalignno{
&\tr_{\al}\equal r_{\al}+\sum_{k>0}(r_{kc+\al}-r_{kc-\al}), \
\al\in R_+,\cr
& y_u\equal \sum_{\tal\in R^a_+} (u,\tal)r_{\tal}\ =\
\sum_{\al\in R_+} (u,\al)\tr_{\al},\ u\in \R^{n}.
&(2.10)
}
$$
If any   pairwise products of these series
are also absolutely convergent, then
$\tr$ is a non-affine $r$-matrix and $[y_u,y_v]\ =\ 0$
for any $u,v\in \R^{n}$.
\hfill\break
\indent
b) Let the  group $W^a$ act in $\f$ by  continuous automorphisms
relative to the norm and $r$ be $W^a$-invariant:
$$
\eqalignno{
& \tw(r_{\tal})\ =\ r_{\tw(\tal)}\for \hbox{all\ } \tw\in W^a,\
\tal\in R^a, &(2.11)
}
$$
for a proper closure of $r$. Then $\tr$ is $W$-invariant and
$$
\align
&s_i (y_u) - y_{s_i(u)} = (u,\al_i)(r_{\al_i}+s_i(r_{\al_i})),
\ 0\le i\le n,\  u\in \R^{n}. \tag 2.12
\endalign
$$

\proof The commutativity in the non-affine case
is proved in [Ch3], Proposition 3.2.
 As to (2.12), see [Ch3], Corollary 3.6 and the
end of Section 1 from [Ch5]. The considerations in the
affine case are the same.  We calculate separately the
sums of the pairwise commutators for any subspaces
$\R \langle \tal, \tbe \rangle \cap R^a $.

Let us fix one more $W^a$-invarint set $\mu\ =\ \{\mu_{\tal},\
\tal \in R^a\}$.

\state
Theorem 2.4.
a) Using the variables $\{x\}$ from  (1.3),(1.5), let
$\f$ be the algebra
$\f^{\flat}$  generated by $\C[W^a]$ and
$$
\C\{x\}\ = \C[\hbox{\rm ct}(x_{\al^{\vee}}+kx_c),\
 \tal=\al+kc\in R^a_+]
$$
with the cross-relations
$\tw x_u = x_{\tw(u)} \tw$, where $\hbox{\rm ct}(t)=(\exp(t)-1)^{-1}$.
Then
$$
\align
&r^{\flat}_{\tal} = \ka_{\tal}\hbox{\rm ct}(x_{a}+kx_c)
(\mu_{\tal}-s_{\tal}),\
 \tal=\al+kc\in R^a, \ a=\al^{\vee}, \tag 2.13
\endalign
$$
is a $W^a$-invariant closed $r$-matrix and $s_i r^{\flat}_{\al_i} +
r^{\flat}_{\al_i} s_i\ =\
\ka_{i}(s_i-\mu)$ for $0\le i\le n$.
\hfill\break
\indent
b) Now  $\f\ =\ \f^{\#}$ is  the algebra
 generated by $\f^{\flat}$ and
$\C\{z\}\ =\ \C[\hbox{\rm ct}(z_{\tal})]$, where
$$z_{u+\kappa c}\ =\
\sum_{i>0} (u, b_i)z_i+ \kappa \xi ,\
u\in \R^n, \for \hbox{ complex\ }
\tz=\{z_1,\ldots,z_n,z_c=\xi\}
$$
commuting with $\C[W^a]$.
  The following functions of $\tz$
$$
\align
&r^{\#}_{\tal}\ =\ \ka_{\tal} \hbox{\rm ct} (z_{\al}+k\xi)
(s_{\tal} - \mu_{\tal}) +r^{\flat}_{\tal}, \ \ \tal=\al+kc\in R^a,
\tag 2.14
\endalign
$$
also form  an  $r$-matrix which is invariant relative to
the diagonal (simultaneous) action $\de$
of $W^a$ on $\{x\}$ and the analogous action $\si$ on $\{z\}$:
$
\si(\tw)(z_u+\kappa\xi)\ = \ z_{\tw(u)}+\kappa\xi.
$
Moreover $\de(s_i)( r^{\#}_{\al_i}) +r^{\#}_{\al_i}\
 =\ 0\ $ for $0\le i\le n$.

\proof The theorem for $\mu=1$ is a straightforward affine
extension of Corollary 3.6 from [Ch3] (see also the end
of Section 2, [Ch5]). These $r$-matrices are  quasi-classical
limits of the quantum $R$-matrices from [Ch4],
Propositions 3.5, 3.8 (cf. (1.6) that is a rational counterpart
of one of them). Calculating the corresponding commutators
(2.1-4) we obtain a set of relations that are the
coefficients of $s_{\tal}$ and  $s_{\tal}s_{\tbe}$
(the latter never coincide with the first). If
$s_{\tal}s_{\tbe}=1$
then $\tal=\tbe$ and the corresponding commutator
equals zero. Hence if the $r$-matrix relations are checked for
one non-zero $\mu$ they are valid for all of them.

\state
Proposition 2.5.
 Let ${\hbox{\it \ae}}\ge \ep>0,\ M>0,\ m\in \Z_+$,
$$
\Xi_{\hbox{\it \ae}}(M)\ =\ \{\ x,x_c,z,\xi, \ \Re (\xi),\Re(x_c)
\ge {\hbox{\it \ae}},\  \hbox{\rm ct} (x_{a}+k x_c),\
\hbox{\rm ct} (z_{\al}+k\xi) <  M\}
$$
 for all $k\in \Z_+,\ \al\in R,\ a=\al^\vee $.
Setting
$f(z,\xi)=\sum_{\tw}f_{\tw}(z,x,\xi,x_c)\tw$ for
 scalar $f_{\tw},\ \tw\in W^a,$
we introduce the norm as follows:
$$
\align
&||f|| \equal
\sum_{\tw}\max\{ | f_{\tw} | \hbox{\ in\ }
\Xi_{\hbox{\it \ae}}(M)\} ||\tw||,\cr
&\where
||\tw||\  = \ \exp ((\hbox{\it \ae}-\ep)l(\tw)(2h-2)^{-1}4^{1-m}),
\tag 2.15
\endalign
$$
 $l(\tw)$ is the length of $\tw\in W^a$ with respect to
the generators $\{s_i,\ 0\le i\le n\}$, $h$  the Coxeter
number, $|\ |$  the absolute value.
 Then  products of any $m$ series from (2.10) are absolutely
convergent (for both $r^{\flat} \and r^{\#}$).
The action of $W^a$ is continuous.

{\it Proof.} Let us start with $\mu=0$.
Without $r^{\flat}, $ (2.15) follows from the estimate
$$
\align
&l(\al+kc)\le kl(a')+\hbox{const} \le k(2h-2)
+\hbox{const},\for k\ge 0,\
a=\al^{\vee},\tag2.16
\endalign
$$
(see e.g. [Ch4],Proposition 1.6 and [Ch7],(1.15)). Here
the factor $4^{1-m}$ is not necessary. Given
$\al(1),\ldots,\al(m)\in R_+,$
let us consider
the  product $\tr^{\flat}_{\al(1)}\cdots
\tr^{\flat}_{\al(m)}$ that is the sum
of
$$
\align
&\Pi_k\ =\ r^{\flat}_{\tal(1)}\cdots r^{\flat}_{\tal(m)}, \
k=\{k(1),\ldots,k(n)\}\subset \Z_+,\ \tal(i)=k(i)\pm \al(i)\in R^a_+.
\endalign
$$
We should fix $C>0$ and calculate the number of the terms
 such that $||\Pi_k||\ >\ C.$  A  certain  problem is that
$\{s_{\tal}\}$ from $\{r^{\flat}\}$
act on the arguments moving them from $\Xi_{\hbox{\it \ae}}(M)$:
$$
\align
&\Pi_k\ =\
\prod_i \ka_{\tal(i)}(\exp x_{\tal^i}-1)^{-1}
\tw, \where \tw=\prod_i(\mu_{\tal(i)}-s_{\tal(i)}),\cr
&\tal^1=\tal(1),\ \tal^2=s_{\tal(1)}(\tal(2)),\ldots,\
\tal^2=(s_{\tal(1)}\cdots s_{\tal(m-1)})(\tal(m)).
\tag2.17
\endalign
$$

\state
Lemma 2.6.
Let $\tal^i=k^i+\al^i,\ \al^i\in R,\quad k_{\pm}=
\max\{0,\pm k^i ,1\le i\le m\},\ K>0$. Then
$$
\align
&c_m k_+
\ge k_-\for c_m=(\nu+1)^{m-1}-1,\tag2.18
\endalign
$$
 where $\nu$ is 1
for $A,D,E$, 3 for $G_2$, and 2 for the other root
systems.
The number of the terms $\Pi_k$ such that $k_+<K$ is less
than $(c_m+1)^{m}K^m$. The length of the corresponding
element $\tw$ is less than $(2h-2)(c_m+1)K$.

\proof We argue by induction on $m$. The inequality
for $k_{\pm}$ is clear for $m=1$ since $k^1=k(1)$
is always non-negative. Supposing that (2.18) is valid for $m$,
let us add one more factor
$\tr^{\flat}_{\tal(0)}$ on the left and denote the new pair of
extreme values of $\{\pm k^i,\ 0\le i\le n\}\ $ by  $k_{\pm}'$.
Then
$$
\align
& k_{+}-\nu k^0\ \le \ k_+'\ \ge \ k^0,\quad
 k_-'\ \le \ k_{-}+\nu k^0, \cr
&c_m(1+\nu)\ k_+'\ge
c_m(k_+'+\nu k^0)\ge k_-\ge
k_-' -\nu k^0\ge k_-' -\nu k_+'.
\endalign
$$
 Hence $(c_m(1+\nu)+\nu)k_+'\ge
k_-'$, which provides the necessary estimate.
As to the length, $l(\tw)=l(\tw^{-1})$,
$\tw^{-1}=s_{\tal^m}\cdots s_{\tal^1}$, and we can use (2.16).

The lemma gives that $||\Pi_k||< \hbox{const} \exp(-K\ep)$
for a rather big $K$ if $k_+>K$.
The number of such terms grows
polynomially in $K$.
If we have a "mixed" product (2.17) where some of $x$ are
replaced by $z$,
then the reasoning is quite similar. We
apply again the induction  taking into consideration mostly
the  first term (with
$k(1)=k^1$). The changes
of the arguments of the others can be controlled in the same way.
When $\mu\neq 0$,  we can use the estimates without $\mu$ for
smaller $m$.
\hfill \llap{\qed}

\state
Corollary 2.7.
Let us denote the operators $y$ from (2.10) considered for
$r^{\flat}$  by $\{y^{\flat}_u, \ u\in \R^n\}$
 and introduce
$$
\align
&x^{\flat}_{u} = y^{\flat}_u+(\rho_{\ka\mu},u),\ x^{\flat}_c =
h_{\ka\mu}'\equal h_{\ka\mu} (\th,\th)/2,
\ x_{u+\kappa c}^{\flat}=x_u^{\flat}+
\kappa x_c^{\flat}. \tag 2.19
\endalign
$$
Then the group algebra
$\C[W^a]$ and $\{x_{\tu}^{\flat}\}$ satisfy  relations
from  Definition 1.1 and form a representation of $\H'$
(which is faithful in  $\H'/(x_c-h_{\ka\mu}')$).
\par
\hfill \llap{\qed}

\sectiontitle
3.  Dunkl operators and KZ
\par

 Let us extend $\C$ -linearly
the standard pairing $(\ , \ )$ to $ \C^n$ and then to
 $\C^{n+2}\ =\ \C^n\oplus \C c \oplus \C d $
 setting
 $ (c,d)=1, \ (c,c)=(c,u)=0=(d,u)=(d,d)$ for $u\in \C^n$
(see e.g. [Ka],Chapter 6).
Given $\tal=\al+kc\in R^a,  \ a \in A,$ the formulas
$$
\eqalignno{
&s_{\tal}(\hu)\ =\  \hu-\{(u,\al)+\nu k\}\al^\vee-
\{\nu k^2(\al^\vee,\al^\vee)/2+(u,\al^\vee)k\} c,\cr
&a'(\hu)\ =\ \hu +\nu a-\{\nu(a,a)/2+(u,a)\}c,
\quad \hu=
 u+\kappa c+\nu d\in \C^{n+2},\cr
& z_{\hu} = \sum_{i=1}^n(u,b_i)z_i+\kappa\xi+\nu\ze,\quad
\si(\tw)(z_{\hu})\equal z_{\{\tw(\hu)\}},\
\si_{\tal}= \si(s_{\tal}).
&(3.1)
}
$$
define   an action  of $ \tw\in W^a$
on $\hu\in \C^{n+2}$ and
$W^a_{\si}\equal \si(W^a)$
 on $z_{\hu}$.

 The linear functions $z_i=z_{\al_i},
 1\le i\le n,\  \xi ,\ze  $ will be regarded
as  coordinates of $ \C^{n+2}$. For instance,
$\partial z_{\tal} /\partial z_i$
 is the
multiplicity of $\alpha _i$ in $\tal=\al +k c \in R^a$,
$\partial z_{\tal} /\partial \xi = k,\
\partial z_{\tal} /\partial \ze = 0$.
 We will also use
the derivatives
$$
\align
&\pa_{\tu}\ =\ \pa_u+\kappa \pa/\pa\ze,\
\partial _{u}(z_{\hv} )=(v ,u), \
\tu=u+\kappa c\in \C^{n+1},\
\hv \in \C^{n+2},
\endalign
$$
with the following evident properties:
$$
\align
&\partial _{r\tu+t\tv}=r\partial _{\tu}
+t\partial _{\tv},\  \si(\tw)(\partial _{\tu})=
\partial _{\tw(\tu)}, \  r,t\in \C,\ \tw\in W^a,\cr
&\partial _{b _i}=\partial /
\partial z_i,\ 1\le i\le n,\ \pa_ c=\pa/\pa\ze.
\tag 3.2
\endalign
$$

We extend $(\rho_{\ka\mu}, \ \cdot\  )$ to a linear function on
$\tu=u+\kappa c \in \C^{n+1}$ by the formulas (see (1.1))
$$
\align
&\trh_{\ka\mu}(\tu) =  (\rho_{\ka\mu}+
h_{\ka\mu}(\th,\th)d/2,\ \tu) =
(\rho_{\ka\mu},u)+ \kappa h'_{\ka\mu},\
h'_{\ka\mu}\equal h_{\ka\mu}(\th,\th)/2,
\tag 3.3
\endalign
$$
 to ensure the relations $\trh_{\ka\mu}(\al_i)=
\ka_i\mu_i(\al_i,\al_i)/2$
\ for all $0\le i\le n$.

Following Theorem 2.4, let us introduce the algebra
$\f^{\#}_\si$
 generated by $\C[W^a_\si]$ and
$\C\{z\}=\C[\hbox{\rm ct}(z_{\tal}),\ \tal\in R^a_+].$
  We will need another   $W^a$ (without $\si$) commuting
with $z$ and the corresponding algebra  $\f^{\#}$ generated
by $W^a$ instead of $W^a_\si$.
 Excluding $\{x\}$, the definition of the sequence of norms
(depending on $m,M$)
 from Proposition 2.5 remains the same.

The algebra
of  differential operators in
$\pa_1,\ldots,\pa_n,\pa_c$
with the coefficients in  $\f^{\#}_{\si}$
will be denoted by $\f^{\#}_\si[\pa]$. We will also
use $\f^{\#}[\pa]$ (the derivatives are always with
respect to $z,\ze$).

\state
Theorem 3.1.
The following family
of differential-difference {\tenbf Dunkl operators}
defined for $\tu=u+\kappa c \in \R^{n+1}$
$$
\eqalignno{
\De _{\tu} \equal
&\pa_u +\kappa\pa/\pa\ze-
\sum_{\tal>0}\ka_{\tal}(u,\al)\hbox{\rm ct}(z_{\tal})
\bigl(\si_{\tal}-\mu_{\tal}\bigr)+ \trh_{\ka\mu}(\tu),\
&(3.4)
}
$$
is commutative.
Moreover, $\{\si_i=\si(s_i)\},\ 0\le i\le n,$ and
$\{\De_{\tu}\}$ satisfy relations (1.4) and the map
$$
\align
&\De:\ s_{\tal}\mapsto \si_{\tal}, \
x_{\tu}\mapsto \De_{\tu}\tag3.5
\endalign
$$
gives an injective homomorphism from $\H'$ into
the algebra   of
convergent series  from
$\f^{\#}_\si[\pa]$. The convergence of differential
operators is coefficient-wise
with respect to the norms for sufficiently big $m,M$.
If $\De_c=\pa/\pa\ze+h'_{\ka\mu}$ is replaced by zero, then $\De$
maps via $\H'_0$.

\proof Without $\{\pa_{\tu}\}$, it follows from
Corollary 2.7. The contribution of the derivatives
is trivial since $[\pa_{\tu}, r^{\flat}_{\tal}]=0$
if $(\tu,\tal)=0$ and
$$
\align
&[\pa_{\tu}, (\tv,\tal) r^{\flat}_{\tal}] -
[\pa_{\tv}, (\tu,\tal) r^{\flat}_{\tal}]  =
[\pa_{(\tv,\tal){\tu}-(\tu,\tal){\tv}}, r^{\flat}_{\tal}] = 0
\hbox{\ for all\ } \tu,\tv.
\endalign
$$

The theorem is valid even when
the map $\si$ satisfies the following weaker properties:
$$
\eqalignno{
& \si_{\tal}z_{\tu}
=z_{\tu'}\si_{\tal},\
\si_{\tal}\pa_{\tu}
=\pa_{\tu'}\si_{\tal},\for
\tu'=s_{\tal}(\tu),\ \tu \in \C^{n+1},
& (3.6)\cr
& \si_{\tal_1}\si_{\tal_2}
= \si_{\tbe_1}\si_{\tbe_2}
\if s_{\tal_1}s_{\tal_2}=s_{\tbe_1}s_{\tbe_2},\
\ \tal, \tbe \in R^a.
& (3.7) \cr
}
$$
Indeed, the necessary relations are written in terms
of commutators (cf. [Ch5], Section 2).

\state
Definition 3.2.
Let us take  $\C[W^a]$ which commutes with
$z,\xi,\ze$ (we  omit $\si$ to
differ it from $\C[W^a_\si]$). Given
$\De \in \f^{\#}_\si[\pa]$, we represent it in the form
$$
\eqalignno{
&\De=\sum_{\tw\in W^a}D_{\tw}\si(\tw),\where  D_{\tw}\hbox{\ are
\ differential},&(3.8)
}
$$
and introduce  the operator  from  $\f^{\#}[\pa]$
$$
\eqalignno{
&\Red(\De)\equal\sum_{\tw\in W^a}D_{\tw}\tw^{-1}
&(3.9)
}
$$
with the coefficients in the completion of the group
algebra $\f^{\#}=\C\{z\}\otimes {}_{\CC}\C[W^a]$.
Replacing $\pa_c=\pa/\pa \ze$ by $-h_{\ka\mu}'$ in
$\Red(\De)$ we obtain $\Red_0(\De) \in
\f^{\#}[\pa_1,\cdots,\pa_n].$
 Both operations are continuous.
\par
\hfill\llap{\qed}

\state
Theorem 3.3.
Given  arbitrary $\De$ and
$W^a_{\si}$-invariant $\De'$ from $\f^{\flat}_\si[\pa],$
$$\Red(\De\De')\ =\ \Red(\De)\Red(\De').$$
If $p\in \C[x_1,\ldots,x_n]^W$, then the  (differential)
{\tenbf OP operators}
$$L_p\equal \Red_0(p(\De_{b_1},\ldots,\De_{b_n})\in
\f^{\#}[\pa_1,\ldots,\pa_n]$$
are pairwise commutative and $W^a_\de$-invariant with respect
to the diagonal action $\de(\tw)=\si(\tw)\otimes\tw $,
where $\tw$ act in $\C[W^a]$ by conjugations (cf. (1.5)).

\proof We completely follow  [Ch5], Theorem 2.4.

\state
Theorem 3.4.
Let us introduce the {\tenbf KZ operators} that
are differential operators
of the first order with convergent coefficients
from $\C\{z\}\otimes_{\CC} \H'$:
$$
\eqalignno{
\D_{\tu} =
&\pa_{\tu} -\sum_{\tal>0}\ka_{\tal}
(\tu,\tal)\hbox{\rm ct}(z_{\tal})
\bigl(s_{\tal}-\mu_{\tal}\bigr)
 + \trh_{\mu\ka}(\tu)- x_{\tu},
\ \tu \in \C^{n+1}.
&(3.10)
}
$$
They are pairwise commutative and satisfy the following
invariance property with respect to the above diagonal action
$\de$ extended to $\H'\supset\C[W^a]$:
$$
\eqalignno{
& \de(\tw)(\D_{\tu})\ =\ \D_{\tw(\tu)},
\ \tw\in W^a,\ \tu\in \C^{n+1}.
&(3.11)
}
$$

\proof First of all, the contribution of the derivations
is zero (see the proof of Theorem 3.1).
Then the commutators
$[\D_{\tu}, \D_{\tv}]$
and the differences
$\tw(\D_{\tu})-\D_{\tw(\tu)}$
for all $\tu,\tv,\tw$ belong to
$\C[W^a]$. We have to check that they vanish.
Theorem 2.4 gives that they really equal zero
in the representation of $\H'$ from
Corollary 2.7. But the latter is faithful when restricred
to $\C[W^a]$.

{\tenbf The isomorphism.\ }
We will show that KZ considered in certain induced representations
of $\H'$ is equivalent to the proper eigenvalue problem for the above
Dunkl operators. It generalizes the constructions from [Ma] and [Ch5].
Let us start with the following general remark.
If $\tw\in W^a$ are boundary operators in a certain algebra
with a norm $\n$,
then the series for $\D\and\De$ and the products of
any $m$ among them
are convergent for rather big
$\Re (\xi)$. Indeed, (2.15) leads to the estimate
$$
\exp(\hbox{\it \ae}(2h-2)^{-1}4^{1-m})>
\max\{\n(s_i),\ 0\le i\le n\}.
$$
It is always so if $W^a$ and $\{x\}$ act in
  finite dimensional representations.

Thus the {\tenbf KZ equation}, which is the system
$$
\align
& \D_{u}\vph(z_1,\ldots,z_n)\ =\ 0,\quad u\in \C^{n},
\tag3.12
\endalign
$$
is well-defined  when the values of $\vph$ are taken in
any finite dimensional representations of $\H'$.
The {\tenbf extended KZ} is obtained for $\tu$ instead of $u$:
 $$
\align
& \D_{u}\phi(z_1,\ldots,z_n,\ze) = 0,\
\pa\phi/\pa\ze +h_{\ka\mu}' = 0,\
h_{\ka\mu}'=h_{\ka\mu}(\th,\th)/2.
\tag3.12a
\endalign
$$
If $\vph$ satisfies (3.12) then $\phi=\vph \exp(-h_{\ka\mu}'\ze)$
is a solution of (3.12a). But this trivial extension is important
for the main theorem below.

We may use standard results about the solutions of
differential equations (assuming that $\Re(\xi)$ is rather
big). Here and further  $\xi$ is considered as a parameter
($\pa/\pa\xi$ does not appear in $\D,\De$).

Following Section 1,
let $V$ be a finite dimensional $\C [W^a]$-module, $\tau$
the corresponding homomorphisms from $\C [W^a]$
to $\ \End V$.
We fix  a set $\la$ = $\{\la_1,..., \la_n\}$
$\subset \C$ and consider the  $\H'_0$-module
$V(\la)$ introduced in (1.17) with the
$\C[W^a]$-homomorphism $tr:\ V(\la)$ $\mapsto V$.
The  homomorhism $\H'_0\mapsto \End V(\la)$
will be denoted by $\hat{\tau}$.

\state
Main Theorem 3.5.
Let $\k$ be the space of solutions $\varphi(z)$
of (3.12) in $V(\la)$ defined in a neighbourhood
of a given point (its dimension coincides with
$\dim V(\la)$ = $|W|\dim V$). Then the map
$tr: \vph\mapsto \psi=tr(\vph)$ is an isomorphism
onto the space $\m$ of solutions of the
{\tenbf quantum many-body problem}
$$
\align
& L_p \psi(z)\ =\ p(\la_1,\cdots,\la_n)\psi(z),\
p(x_1,\cdots,x_n)\in \C[x]^W,
\tag3.13
\endalign
$$
for the operators $\{L\}$ introduced in
Theorem 3.3.

{\it Proof.} The statement is a direct generalization
of Theorem 4.6 from [Ch5].
We will remind the main steps of the proof
(adopted to the affine case).

In the set up of Theorem 2.5, let us pick a set
$ Z\subset \C^{n+1}$ obtained from
$\Xi_{\hbox{\it \ae}}(M)$ by certain cuts off
and obeying the following conditions.
It is connected and simply connected. The image of
the intersection
$\ \bigcap_{\tw} \tw(Z)\ $
 in the quotient
$\Xi_{\hbox{\it \ae}}(M)/W^a$ is connected.
Assuming that  $\Re (\xi)$ is rather big,
 we can fix an invertible analitical
 solution $\Phi(z,\ze)$ of (3.12a) for $z\in Z$
and arbitrary $\ze$
 with the values in $\End V(\la)$.

The functions
$\si(\tw)\Phi$,
$\tw\in W^a$,
are well-defined in  open subsets of $Z$,
we may introduce the "monodromy matrices" $T$:
$$
\align
&\hat{\tau}(\tw)\Phi(z,\ze)\ = \
\si(\tw^{-1})\bigl(\Phi(z,\ze)\bigr)T_{\tw}(z,\ze),\quad
\tw\in W^a,
\tag 3.14
\endalign
$$
which are well-defined for almost all
$z\in Z$ and  locally constant (use the invariance of
$\D_{\tu}$). They satisfy
the  one-cocycle relation
$$
T_{\tw_1\tw_2}=\si(\tw_2^{-1})\bigl(T_{\tw_1}\bigr)T_{\tw_2},
\quad
\tw_1,\tw_2\in W^a,
$$
which results in the following  action $\bar{\si}$ of $W^a$:
$$
\align
&\bar{\si}(\tw)(F(z,\ze))\equal
\si(\tw)\bigl(F(z,\ze)\bigr)T_{\tw^{-1}}(z,\ze),\quad
\tw\in W^a,
\tag 3.15
\endalign
$$
on  $\End V(\la)$-valued functions $F$ defined
for almost all $z\in Z$.

Substituting $\bar{\si}_\al=\bar{\si}(s_\al) \for
\hat{\tau}(s_\al)$
we  rewrite KZ for $\Phi$
as the system
$$
\align
&\bar{\De}_{\tu}(\Phi)\ =\ \hat{\tau}(x_{\tu})\Phi,\
\tu\in \C^{n+1},
\tag 3.16
\endalign
$$
where the operators $\bar{\De}_{\tu}$ are introduced
by  formulas (3.4) with $\si$  replaced
by $\bar{\si}$.  The latter obeys relations (3.6), (3.7),
which ensure the validity of Theorems 3.1,3.3.
The operators $\bar{L}_p$ constructed for $\bar{\si}$
(by replacing $\bar{\si}({\tw})$ on the right with $\tw^{-1}$)
 coincide with $L_p$ for $\si$.
Hence,
$$
\align
& p(\bar{\De}_{\tu})(\Phi)\ =\ (\hat{\tau}(p(x_{\tu}))(\Phi)\
=\ p(\la)\Phi,\cr
& L_p (\Phi)\ =\ p(\la)\Phi \for p(x_1,\cdots,x_n)\in \C[x]^W.
\tag3.17
\endalign
$$

The last formula contains no $\{x\}$ and  therefore
commutes with $tr$. More precisely, given $e\in V(\la),$
$$ tr(L_p (\Phi e))\ =\ p(\la)tr(\Phi e).$$
Since an arbitrary solution  $\vph\in\k$ can be represented
in the form $\ \Phi e\ $ for a proper $e$,  the
image of $\k$ belongs to $\m$. The dimension of the latter is
not more than $\dim \k$. However $tr$ has no
kernel  due to Proposition 1.4 (as it was checked  in [Ch5]).
\par
\hfill\llap\qed

It is worth mentioning that one can introduce the monodromy
of KZ more traditionally. It is necessary to fix a point
$z^0$ and to replace $\Phi$ in right-hand side of (3.14)
by its analitical continuation along a certain path from
$z^0$ to $\tw(z^0)$ (see [Ch2,5]). This approach gives
a representaion
of the "elliptic" braid group
which is directly connected with the induced representations
of the double affine Hecke algebras from [Ch7,8].

\sectiontitle
4. Examples

We will calculate the first (quadratic) $L$-operators
for the simplest  $\mu\subset \{\pm1\}$, $ \mu=0$ and
discuss their basic
properties. More complete analysis will be continued in
the next paper(s).

The following elliptic functions $\vsi, \ \vth$
"almost" coincide (but do not coincide)
with the classical $ \ze,\ \vartheta_1$.
To avoid confusions we changed a little the standard
notations.
Let
$$
\align
&\vsi(t)\ =\ \sum_{k=0}^{\infty} \hbox{\rm ct}(k\xi+t)-
\sum_{k=1}^{\infty} \hbox{\rm ct}(k\xi-t),\cr
&\vth(t) = (\exp(t/2)-\exp(-t/2)) \prod_{k=1}^{\infty}
(1-\exp(-k\xi+t))(1-\exp(-k\xi-t)),\cr
&\vrh(t)\ =\ \sum_{k=1}^{\infty} k\bigl(
 \hbox{\rm ct}(k\xi+t)+\hbox{\rm ct}(k\xi-t)\bigr).
\tag 4.1
\endalign
$$
Here $t,\xi\in \C,\ \Re(\xi)>0.$
All these functions are $2\pi\imath\Z$-invariant.
One has the following relations (which can be deduced
from the corresponding properties of $\ze  \and \vth_1$
or proved directly):
$$
\align
&\vsi(t + m\xi)=\vsi(t )+m,\quad
\vth(t +m\xi)=-\exp(t +\xi/2)\vth(t ),\cr
&\vsi(t )+\vsi(-t )=-1,\
\vth(-t )=-\vth(t ),\ \vrh(-t )=\vrh(t ),\cr
&\pa(\log\vth(t ))/\pa t = \tvs(t )\equal \vsi(t )+1/2,\
\pa(\log\vth(t ))/\pa \xi = \vrh(t ),\cr
& \vrh(t -\xi)=\vrh(t )+\vsi(t),\
\vsi'\equal\pa\vsi/\pa t =\vpi-\tvs(t )^2-2\vrh(t ).
\tag 4.2
\endalign
$$
As to the latter (up to a constant $\vpi$), check that the
difference of the two functions has no poles
and is periodic with respect to the shifts by $\xi$ (everything
is periodic relative to $2\pi\imath \Z$).

Let us take $\mu=\pm 1$ and the corresponding
 one-dimensional $V=\C_\mu$ (see (1.8)). Our first aim is to
determine $L_2=L_{p_2}$ (Theorem 3.3) for
$$p_2(x_1,\ldots,x_n)\ =\ \sum_{i=1}^n x_ix_{\al_i},\
x_{\al_i}=\sum_j (b_j,b_i)x_j.$$
The calculations are rather simple because
$\Red_0(\si_{\tal}-\mu_{\tal})=0$ :
$$
\align
&\Red_0(\De_{b_i}\De_{\al_i})\ =\
\pa_i\pa_{\al_i}+(\rho_{\ka\mu},\al_i)\pa_i+
(\rho_{\ka\mu},b_i)\pa_{\al_i}+\cr
&(\rho_{\ka\mu},\al_i)(\rho_{\ka\mu},b_i)+
\Red_0\{-\sum_{\tal>0}\ka_{\tal}(b_i,\tal)\ct(z_{\tal})
(\si_{\tal}-\mu_{\tal})\pa_{\al_i}\},
\tag 4.3
\endalign
$$
where the last term equals $$+\sum_{\al+kc>0}
\ka_{\al}\mu_{\al}(b_i,\al)(\al_i,\al^{\vee})
(\ct(z_{\al}+k\xi))
(\pa_{\al_i}-kh_{\ka\mu}').
$$
Here we applied (3.1),(3.2) and replaced
$\pa/\pa\ze$ by $-h_{\ka\mu}'$.
To sum up the terms (4.3) with respect to $i$,
we use the definition of $\rho_{\ka\mu}$ and
the relations
$$b\ =\ \sum_{i=1}^n (b,b_i)\al_i\ =
\ \sum_{i=1}^n (b,\al_i)b_i.$$
Finally, $L_2 \ = \
\Red_0(\sum_{i=1}^n\De_{b_i}\De_{\al_i})\ =$
$$
\align
&\sum_{i=1}^n\pa_i\pa_{\al_i}+
2\pa_{\rho_{\ka\mu}}+(\rho_{\ka\mu},\rho_{\ka\mu})+
 2\sum_{\al\in R_+}\ka_{\al}\mu_{\al}
\bigl(\vsi(z_{\al})\pa_{\al}-
h_{\ka\mu}'\vrh(z_\al)\bigr) =\cr
&\sum_{i=1}^n\pa_i\pa_{\al_i}+
(\rho_{\ka\mu},\rho_{\ka\mu})+
 2\sum_{\al\in R_+}
\ka_{\al}\mu_{\al}\bigl(\tvs(z_{\al})\pa_{\al}-
h_{\ka\mu}'\vrh(z_\al)\bigr).
\tag 4.4
\endalign
$$

The next calculation will be a reduction of $L_2$ to
the Schr\"odinger  operator (without linear differentiations).
We will introduce the following elliptic generalization
of the "standard product"  playing the
main  role in the Macdonald theory, Heckman-Opdam theory,
 and the theory of
integral solutions of KZ :
$$\om(z)=\om(-z)\equal \prod_{\al\in R}
\vth(z_{\al})^{
\ka_{\al}\mu_{\al}/2}.$$
Actually we will need in this paper only the formulas (see (4.2):
$$
\align
&\pa_{u}(\om)\ =\ \om\ \sum_{\al\in R_+}
\ka_{\al}\mu_{\al}(u,\al_i)\tvs(z_{\al}),\for
u\in\C^n,\cr
&\pa\om/\pa\xi\ =\ \om\ \sum_{\al\in R_+}
\ka_{\al}\mu_{\al}\vrh(z_{\al}).
\tag4.5
\endalign
$$
The first gives that $H_2\equal \om L_2 \om^{-1}$ is
free of linear differential operators. More precisely,
$H_2=\sum_{i=1}^n\pa_i\pa_{\al_i}+
(\rho_{\ka\mu},\rho_{\ka\mu})- U(z),\where $
$$
\align
U(z)\ =\
 2\sum_{\al\in R_+}
\ka_{\al}\mu_{\al} \bigl(&\tvs(z_{\al})
\pa_{\al}(\om)\om^{-1}+
h_{\ka\mu}'\vrh(z_\al)\bigr)+\cr
&\sum_{i=1}^n\bigl((\pa_{\al_i}(\om)\om^{-1})
(\pa_{i}(\om)\om^{-1})-
\pa_i\{\pa_{\al_i}(\om)\om^{-1}\}\bigr)=\cr
\sum_{\al>0}
\ka_{\al}\mu_{\al}\bigl((\al,\al)&\vsi'(z_{\al})+
2h_{\ka\mu}'\vrh(z_\al)\bigr)+\cr
&\sum_{\al,\be>0}
\ka_{\al}\mu_{\al}
\ka_{\be}\mu_{\be}(\al,\be)\tvs(z_\al)\tvs(z_\be).
\tag4.6
\endalign
$$

\state
Lemma 4.1.
$$
\align
&\sum_{\al,\be>0}
\ka_{\al}\ka_{\be}(\al,\be)\tvs(z_\al)\tvs(z_\be)=
h_{\ka}'\sum_{\al>0}\ka_{\al}\tvs^2(z_\al)+C(\ka).
\tag4.7
\endalign
$$

\proof
Let us fix $b\in B$ and replace $z_u$ by $z_{b'(u)}=z_{u-(b,u)c}=
z_u-(b.u)\xi$ for $u=\al,\be$ in (4.7).
The change of the left-hand side is
$$
\align
&\sum_{\al,\be>0}
\ka_{\al}\ka_{\be}(\al,\be)\bigl((b,\al)\tvs(z_\al)+
(b,\be)\tvs(z_\be)+(b,\al)(b,\be)\bigr)=\cr
&h_{\ka}'\sum_{\al>0} 2\ka_{\al}(b,\al)\tvs(z_\al)+
(h'_{\ka})^2(b,b).
\tag4.8
\endalign
$$
Here we used the main property of $h'_{\ka}$:
$$\sum_{\al>0}\ka_\al(u,\al)(v,\al)\ =\ h'_{\ka}(u,v)
\for u,v\in \C^n.$$
The same  holds for the right-hand side. Hence, their
 difference is $B$-periodic and
has no singularities. The latter can be checked directly or deduced
from  (4.8) with $t^{-1}$ instead of $\tvs(t)$ (use
 the $r$-matrix relations).
Thus the difference is a  constant $C$
depending on $\ka$.

Finally, applying the lemma and
replacing $2\vrh(z_\al)+\tvs(z_\al)^2$ by
$\varpi-\vsi'(z_\al)$ (see (4.2)), we arrived at
the formula for $U$ and the following

\state
Theorem 4.2.
a) If $\mu\subset \{\pm1\}$ and $V=\C_{\mu}$ is the corresponding
one-dimensional representation of $W^a$, then the reduction
procedure for $p_2=\sum_i x_ix_{\al_i}$ gives the operator
$L_2$ conjugated (by $\om$) with
$$
\align
&H_2\ =\ \sum_{i=1}^n\pa_i\pa_{\al_i}+
\sum_{\al>0}\ka_{\al}\mu_{\al}
\{h'_{\ka\mu}-(\al,\al)\}\vsi'(z_\al)+\cr
&(\rho_{\ka\mu},\rho_{\ka\mu})-
\varpi h'_{\ka\mu} \sum_{\al>0} \ka_{\al}\mu_{\al} -C(\ka\mu).
\tag4.9
\endalign
$$
\hfill\break
\indent
b) The operator $H_2$  can be included
into the family of pairwise commutative
differential operators $H_p\equal\om L_p\om^{-1},
\ p\in \C[x]^W$, which are $W$-invariant. Their coefficients
are $B$-periodic
with respect to the action
$z_u\to z_u-(u,b)\xi,\ b\in B$. They are
self-adjoint relative to the complex involution
taking $z_u$ to $-z_u$  and leaving $ \pa_u$
invariant.
\hfill\break
\indent
c) Operators $\{L_p\}$ are W-invariant as well. Moreover,
they are $B$-invariant for the  action:
$$z_u \to z_u-(u,b)\xi,\quad \pa_u \to \pa_u+(u,b)h_{\ka\mu},\ \
 b\in B,\ u\in \C^n, $$
and formally self-adjoint with respect to the following
paring:
$$\langle f(z),g(z)\rangle \ =\ \int \om^2
f(z)g(-z)dz_1\ldots dz_n.$$

\proof  The previous calculation gives a). Begining
with  c),
the invariance relative to $W^a$ (generated by
$W$ and $A$) is due to Theorem 3.3. It can be naturally extended
to the the bigger group with $B$ instead of $A$. We will not
discuss this extension in this paper. The self-adjointness
results from the same property of $\De_{\tu}$,  which
can be checked directly using the definition of $\om$.
It gives the analogous properties of $H$. For instance,
let us  check the  periodicity :
$$
\align
&\tilde{\omega}^{-1}\pa_u \tilde{\omega}=
\pa_u+\sum_{\al>0}\ka_{\al}\mu_{\al}(b,\al)(\al,u)=
\pa_u+ h_{\ka\mu}'(u,b),\cr
&\where \tilde{\omega}=\omega (z_\al\to z_\al-(b,\al)\xi)=
\om \ \exp(-\sum_{\al\in R}\ka_\al\mu_\al(b,\al)\xi/2).
\endalign
$$

Without going into detail we mention that one  can
generalize the construction of the shift operators
from [Op,He] to the elliptic case. It is directly
connected with  Theorem 3.5 for $\C_\mu$
(see [FV]). The most interesting applications
of these operators are expected when  $\mu=1$ because in
this case the operators
$L_p$ preserve certain  subspaces of
$W$-invarint elliptic functions.

To define these spaces let us fix $m\in \Z_+$ and
introduce  the  set
$$
\align
&\tP^+_m\equal \{ \tbe = k_1\om_1+\ldots+k_n\om_n +k c, \
(\tbe,\th)\le m' \}, \ m'\equal m(\th,\th)/2,\cr
&\where \om_i=(\al_i,\al_i)b_i/2,\quad k_1,\ldots,k_n\in
\Z_+, \ k\in \Z.
\tag 4.10
\endalign
$$
The linear space generated by the orbitsums
$$
\align
&\Upsilon_{\tbe_+} = \sum_{\tbe\in W^a(\tbe_+)}
\exp(z_{\tbe }+m'\ze)\for
\tbe_+\in \tP^+_m
\tag 4.11
\endalign
$$
over the algebra of formal series
$\sum_{l<l_0} c_l\exp(l\xi),\ c_l\in \C$, will be denoted
by $\l_m$. This construction is due to Looijenga and
closely related to the characters of Kac-Moody algebras.
The operators $\{L_p\}$ for $\mu=1$
leave  $\l_m$ invariant if $m'=-h'_{\ka}$. Moreover they
preserve subspaces $\l_m(\tbe_+)\for \tbe_+\in \tP^+_m$
generated by
$$
\align
& \Upsilon_{\tga_+}\hbox{\ \ such that\ \ } \tga_+ = \tbe_+ -
\sum_{i=0}^n k_i\al_i\in \tP^+_m,\ \{k_i\}\subset \Z_+.
\endalign
$$
It results directly from the corresponding properties
of the elliptic Dunkl operators and allows us
to introduce the elliptic
 Jacobi-Jack-Macdonald polynomials
$J_{\tbe_+}$ as  eigenfunctions of $\{L_p\}$ in
$\l_m(\tbe_+)$  with leading terms $\Upsilon_{\tbe_+}$.
A further discussion will be continued in the
next papers.

{\tenbf Parabolic operator}. A  demerit of
the above constructions is the constraint
$\pa/\pa\ze+h_{\ka\mu}'=0$ corresponding to
the condition $x_c=0$ in the Hecke
algebra $\H'$.
We will show that something can be
done even without this restriction.

Let  $\De_{\hu} = \De_{\tu}+\nu\De_d$ for $\hu=\tu+\nu d\in
\C^{n+2},$
$$
\align
&\De_d\ =\ \pa/\pa\xi - \sum_{\al\in R}\sum_{k\in \Z_+}
\ka_{\al}k\ct(z_{\al}+k\xi)(\si_{\al+k\xi} -\mu_{\al}).
\tag4.12
\endalign
$$
The operators $\De_{\hu}$ are not pairwise commutative
but still satisfy the following cross-relations (see (1.4)):
$$
\align
&\si_i\De_{\hu}-\De_{\{s_i(\hu)\}}\si_i\ =\ \ka_i(\hu,\al_i ),
\quad 0\le i\le n,\ \hu\in \C^{n+2},
\tag 4.13
\endalign
$$
relative to the action from (3.1).
It gives (together with the previous considerations)
the following theorem.

\state
Theorem 4.3.
The operator $\ \m = 2\De_d\De_c +
\sum_{i=0}^n\De_{b_i}\De_{\al_i}\ $
 and  its reduction  $M  = \Red(\m)$
are $W^a$-invariant.
If $V=\C_\mu,\ \mu=1,\ \pa/\pa\ze = m',\ m\in \Z_+,$  then
$$
\align
M\ =\
&2(m'+h'_{\ka})\pa/\pa\xi + \sum_{i=1}^n\pa_i\pa_{\al_i}+
(\rho_{\ka\mu},\rho_{\ka\mu})+\cr
& 2\sum_{\al\in R_+}
\ka_{\al}\mu_{\al}\bigl(\tvs(z_{\al})\pa_{\al} +
m'\vrh(z_\al)\bigr),\cr
 N\equal &\om M\om^{-1}\ =\  2(m'+h'_{\ka})\pa/\pa\xi + H_2 \hbox{\ (see
(4.9))}.
\tag 4.14
\endalign
$$
The operator $M$ preserves the spaces $\l_m$ and $\l_m(\tbe_+)$ for
arbitrary  $\tbe_+\in \tP^+_m$.
\par
\hfill\llap{\qed}

The operator $N$ was introduced by Etingof and Kirillov
in [EK] for ${\goth {sl}}_n$ together with its certain eigenfunctions
(the generalized characters that are the traces of
proper vertex operators of $\hat{{\goth {sl}}_n}$). In a recent work, they
extended the definition of $M, N$  to arbitrary root systems
and proved directly the properties mentioned in the
theorem. To be more precise, their formulas are different but
with certain minor changes seem to be equivalent to (4.14)
(e.g. they use more special parameters).
If it is so, then our approach
(based on the Dunkl operators) gives
another proof of their result. The construction of
the generalized characters is still known for ${\goth {sl}}_n$
only.

{\bf Matrix Schr\"odinger operator.}
The next application (which is a straight forward
extension of Corollary 2.8 from [Ch5])
 will be for
arbitrary representations and $\mu=0$.
Let us calculate $L_2^0=L_2^{\mu=0}$
for $p=p_2$ (see above). Applying $\Red$ and
imposing the condition $\pa/\pa\ze=0$,
one has:
$$
\align
L_2^0\ =&\ \sum\limits_{i=1}^n\pa_{b_i}\pa_{\al_i}-
\sum\limits_{\tal=\al+kc>0} \ka_\al (\al,\al)
 \ct'(z_{\tal})s_{\tal}+\cr
&\sum_{\tal,\tbe>0}\ka_\al \ka_\be (\al,\be)
\ct(z_{\tal})\ct(\si_{\tal}(z_{\tbe}))s_{\tbe}s_{\tal},
\tag 4.15
\endalign
$$
where $\ct'(t)=\pa\ct(t)/\pa t=-(\exp(t/2)-\exp(-t/2))^{-2}$.
Following [Ch5], Lemma 2.7, we check that the contribution
of the terms with $\tal\neq\tbe$ in the last sum
equals zero. Hence we arrived at the following theorem:

\state
Theorem 4.4.
The differential $\C[W^a]$-valued operators
$$
\eqalignno{
&L_2^0\ =\ \sum\limits_{i=1}^n\pa_{\be_i}\pa_{\al_i}+
\sum\limits_{\tal>0}(\al, \al)\ka_\al \ct'(z_{\tal})
(\ka_\al-s_{\tal})
&(4.16)
}
$$
and $L_p^{\mu=0}$ defined for $p\in \C[x]^W$
are pairwise commutative. Moreover they are
$W^a$-invariant with respect to the $\de$-action
on $z$ and on $\C[W^a]$ (by conjugations).
When considered in finite dimensional representations
of the latter,
the coefficients are convergent matrix-valued
functions for sufficiently big $\Re(\xi)$.
\par
\hfill\llap{\qed}

We can obtain the  scalar
OP operators (for arbitrary root systems)
from this construction as well. Let
 $\{s_{\tal}\}$ be taken
in one-dimensional representations $\C_{\vep}$ (see
(1.8)). Then
$$
\eqalignno{
&L_2^0\ =\ \sum\limits_{i=1}^n\pa_{\be_i}\pa_{\al_i}+
\sum\limits_{\al>0}(\al, \al)
\ka_\al (\ka_\al-\vep_\al)
\vsi'(z_\al).
&(4.17)
}
$$

The corresponding $L_p$ are $W$-invariant and their
coefficients are elliptic = $B$-periodic (cf. Theorem 4.2).
Generally speaking, the coefficients are "matrix"
elliptic functions with the values in the endomorphisms
of vector bundles over elliptic curves.
Ignoring the differential operators in (4.16) and
substituting "good"  $z$, we obtain
"periodic" generalizations of Haldane-Shastry
hamiltonians. Presumably the points of finite
order of the corresponding elliptic curve and
the critical points of the scalar hamiltonians
((4.17) without the differentiations and
after a proper normalization)
lead to integrable models (cf. [BGHP],[F]).

\bigskip
{\bf References}
\medskip
\ninerm
\baselineskip=11pt

\item{[B]}
Bourbaki, N.: Groupes et alg\`ebres de Lie. Ch. 4--6.
Hermann, Paris 1969.

\item{[BGHP]}
Bernard, D., Gaudin M., Haldane F.D.M., Pasquier V.:
Yang-Baxter equation in spin chains with long range interactions.
Preprint SPhT-93-006 (1993).

\item{[Ch1]}
Cherednik, I.V.: Generalized braid groups and local $r$-matrix
systems. Doklady Akad. Nauk SSSR, {\bf 307}:1, 27--34 (1989).

\item{[Ch2]}
Cherednik, I.: Monodromy representations for generalized
Knizhnik--Zamolodchikov equations and Hecke algebras.
Preprint ITP-89-74E (Kiev, 1989).
Publ. of RIMS. {\bf 27}:5, 711--726 (1991).

\item{[Ch3]}
Cherednik, I.: A unification of Knizhnik--Zamolodchikov
and Dunkl operators via affine Hecke algebras. Preprint RIMS--{\bf 724}.
Inventiones Math. {\bf 106}:2, 411--432 (1991).

\item{[Ch4]}
Cherednik, I.: Quantum Knizhnik- Za\-mo\-lod\-chi\-kov
equa\-tions and affine
root systems. Commun. Math. Phys. {\bf 150}, 109--136 (1992).

\item{[Ch5]}
Cherednik, I.: Integration of Quantum many-body problems by affine
Knizhnik-Za\-mo\-lod\-chi\-kov equations. Pre\-print RIMS--{\bf 776} (1991),
to be published in Advances in Math.

\item{[Ch6]}
 Cherednik, I.: The Macdonald constant term conjecture.
IMRN  {\bf 6 }, 165--177 (1993).

\item{[Ch7]}
Cherednik, I.: Induced representations of double affine Hecke
algebras and applications. Preprint MPI /93--76 (1993),
to be published in MRL.

\item{[Ch8]}
Cherednik, I.: Double affine Hecke algebras,
Knizhnik- Za\-mo\-lod\-chi\-kov equa\-tions, and Mac\-do\-nald's
ope\-ra\-tors.
IMRN  {\bf 9}, 171--180 (1992).

\item{[E]}
Etingof, P.I.: Representations of affine Lie algebras,
elliptic $r$-matrix systems, and special functions.
Hep-th bull. board 9303018 (1993), to be published in Commun. Math.
Phys.

\item{[EK]}
Etingof, P.I., Kirillov Jr., A.A.: Representations of affine Lie algebras,
parabolic equations, and Lam\'e functions.
Hep-th bull. board 9310083 (1993), submitted to Duke Math. J.

\item{[F]}
Frahm H.: Spectrum of a spin chain with inverse square exchange.
Preprint (1993).

\item{[FV]}
Felder, G.,Veselov, A.P.: Shift operators for the quantum
Calogero - Sutherland problems via Knizhnik - Zamolodchikov
equation. Communs Math. Phys. {\bf 160}, 259--274 (1994).

\item{[FW]}
Felder, G., Wieczerkowski, C.: The Knizhnik - Zamolodchikov -
Bernard equation on the torus. Preprint 1993, to appear in
Proceedings of the Vancower Summer School on Mathematical
Quantum Field Theory (August 1993).

\item{[He]}
Heckman G.J.: An elementary approach to the hypergeometric shift
operator of Opdam. Invent. math. {\bf 103}, 341--350 (1991).

\item{[Ka]}
Kac V.: Infinite dimensional Lie algebras. Birkh\"auser,
Boston,Basel,Stuttgart, 1983.

\item{[L]}
Lusztig G.: Affine Hecke algebras and their graded version.
J. of the AMS {\bf 2}:3, 599--685 (1989).

\item{[M]}
Macdonald, I.G.: A new class of symmetric functions.
Publ.I.R.M.A., Strasbourg, Actes 20-e Seminaire Lotharingen.
 131--171 (1988).

\item{[Ma]}
Matsuo A.: Knizhnik-Zamolodchikov type equations and zonal
spherical functions. Preprint RIMS--{\bf 750} (Ap., 1991).

\item{[OOS]}
Ochiai H., Oshima T., Sekiguchi, H. : Commuting families
of symmetric differential operators. Preprint (1994).

\item{[OP]}
Olshanetsky, M.A., Perelomov, A.M.: Quantum integrable systems
related to Lie algebras. Phys. Rep. {\bf 94}, 313--404 (1983).

\item{[O]}
Opdam, E.M.:  Some applications of hypergeometric shift
operators. Invent.Math.{\bf 98},  1--18 (1989).

\vskip 10pt

\font\cc=cmcsc8 

{\cc IVAN CHEREDNIK : 

Dept. of Math., CB \#3250 Phillips Hall,

University of North Carolina at Chapel Hill,

Chapel Hill, N.C. 27599-3250, USA}

{\eightrm internet:  chered\@math.unc.edu}

\end